\documentclass[global]{svjour}
\newcommand{\defn}{\emph}      % First/defining use of a phrase.
\newcommand{\foreign}{\textit} % Foreign phrase.
\newcommand{\latin}{\foreign}  % Latin phrase.

\newcommand{\ie}  {\latin{i.e.}}   % in other words
\newcommand{\etal}{\latin{et al.}} % and all
\newcommand{\eg}  {\latin{e.g.}}   % for example
\newcommand{\etc} {\latin{etc.}}   % et cetera (and so on)

\newcommand{\const}[1]{{\rm #1}}  % Mathematical constant.
\newcommand{\e}{\const{e}}        % Base of natural logarithm.
\newcommand{\imag}{\const{i}}     % Imaginary unit.
\newcommand{\diff}{\mathrm{d}}    % Differential operator.
\newcommand{\ket}[1]{|#1\rangle}        % Dirac ket (vector)
\newcommand{\bra}[1]{\langle #1|}       % Dirac bra (covector)
\newcommand{\E}[2]{\bra{#2}#1\ket{#2}}  % Expectation value of operator #1 applied to vector #2.
\newcommand{\eff}{\mathcal{F}}          % Computational effort.
\newcommand{\dif}{\mathcal{D}}          % Computational difficulty.
\usepackage{amssymb}
\usepackage{times}
\usepackage{psfig}

\journalname{Quantum Information Processing}
\title{On the Interpretation of Energy as the Rate of Quantum Computation}
\author{Michael P.~Frank}
\date{\today}
\institute{FAMU-FSU College of Engineering, Dept. of Electrical \&
Computer Engineering, 2525 Pottsdamer St., Rm. 341, Tallahassee, FL
32310, \email{mpf@eng.fsu.edu}.}

\begin{document}
\maketitle

\begin{abstract}
Over the last few decades, developments in the physical limits of
computing and quantum computing have increasingly taught us that it
can be helpful to think about physics itself in computational terms.
For example, work over the last decade has shown that the energy of
a quantum system limits the rate at which it can perform significant
computational operations, and suggests that we might validly
interpret energy as in fact {\em being} the speed at which a
physical system is ``computing,'' in some appropriate sense of the
word. In this paper, we explore the precise nature of this
connection. Elementary results in quantum theory show that the
Hamiltonian energy of any quantum system corresponds exactly to the
angular velocity of state-vector rotation (defined in a certain
natural way) in Hilbert space, and also to the rate at which the
state-vector's components (in any basis) sweep out area in the
complex plane.  The total angle traversed (or area swept out)
corresponds to the action of the Hamiltonian operator along the
trajectory, and we can also consider it to be a measure of the
``amount of computational effort exerted'' by the system, or {\em
effort} for short.  For any specific quantum or classical
computational operation, we can (at least in principle) calculate
its {\em difficulty}, defined as the minimum effort required to
perform that operation on a worst-case input state, and this in turn
determines the minimum time required for quantum systems to carry
out that operation on worst-case input states of a given energy. As
examples, we calculate the difficulty of some basic 1-bit and
$n$-bit quantum and classical operations in an simple unconstrained
scenario. 
\end{abstract}
%\end{opening}

\newpage
\tableofcontents

\section{Introduction}
Over the years, the quest to characterize the fundamental physical
limits of information processing has also helped to give us a deeper
understanding of physics itself.  For example, Shannon's studies of
the limits of communication \cite{Shannon-48} taught us that the
entropy of a system can also be considered to be a measure of the
expected amount of unknown or incompressible information that is
encoded in the state of that system.  Landauer's \cite{Landauer-61}
and Bennett's \cite{Bennett-73} analyses of the lower limit to the
energy dissipation of computational operations led to Bennett's
resolution \cite{Bennett-87} of the famous Maxwell's demon paradox,
via the realization that the demon's record of its past perceptions
is a form of physical entropy, which must be returned to the
environment when that information is erased.  More recently,
Margolus and Levitin \cite{Margolus-Levitin-98} showed that the
energy of a quantum system limits the rate at which it can perform
computational ``operations'' of a certain type, namely, transitions
between distinguishable (orthogonal) quantum states.  In the last
few years, several articles by Lloyd and colleagues
\cite{Lloyd-00,Lloyd-02,Lloyd-Ng-04} have elaborated on this theme
by suggesting that we can think of all variety of physical systems
(ranging from particles and black holes to the entire universe) as
comprising natural computers, with each system's ``memory capacity''
given by its maximum entropy, and its ``computational performance''
given by its total energy.  We should also note that Ed Fredkin has
been promoting a universe-as-computer philosophy for many decades.

The concept of interpreting physics as computing is certainly an
exciting theme to pursue, due to its promise of conceptual
unification, but we would like to proceed carefully with this
program, and take the time to understand the details of this
potential unification thoroughly and rigorously. While taking care
to get all of the details exactly right, we would like not only to
establish that a given physical quantity ``limits'' or ``relates
to'' a given informational or computational quantity, but also
justify the even stronger statement that the physical quantity
actually {\em is}, at root, a fundamentally informational or
computational quantity, one that has been traditionally expressed in
terms of operationally defined physical units for reasons that can
be viewed as being merely historical in nature.

As one the most famous examples of this type of conceptual
progression, Rudolph Clausius \cite{Clausius-1865} first defined
(differential) entropy as the ratio of differential heat to
temperature, $\diff S = \diff Q/T$, and at the time, entropy had no
further explanation. Later, Ludwig Boltzmann \cite{Boltzmann-1872}
proposed the relation $S\propto -H = \int f\log f\,\diff\vec{\xi}$
(where $f$ is a probability density function ranging over particle
energies or velocity vectors $\vec{\xi}$), which was backed up by
his ``H-theorem'' showing that $H$ spontaneously decreases over time
for statistical reasons.  In subsequent decades, this relation for
entropy evolved and was generalized to become Boltzmann's eventual
epitaph $S = k\log W$, which related entropy to the logarithm of the
number of ways $W$ of arranging a system
\cite{Cercignani-98}.\footnote{The references to Clausius and
Boltzmann in this paragraph are also taken from
\cite{Cercignani-98}.} Boltzmann's logarithmic quantity $H$ (in a
discrete and negated form) was later recognized by Shannon and
others to also be an appropriate measure of the information content
of a system.  But, Boltzmann's fundamental insight regarding the
nature of entropy can be viewed as having gone far beyond just
\emph{relating} a physical quantity to an information-based one.
Rather, it can be viewed as telling us that physical entropy, at
root, \emph{is} really nothing but an informational quantity, one
which merely manifests itself in terms of measurable physical units
of heat and temperature due to the fact that these quantities
themselves have an origin that is ultimately of a statistical
nature, \eg, heat as disorganized energy.

Indeed, the long-term quest of physics to eventually create a grand
unified ``theory of everything'' can be viewed as the effort to
eventually reveal {\em all} physical concepts, quantities, and
phenomena as being manifestations of underlying structures and
processes that are purely mathematical and/or statistical in nature,
and that therefore have an informational/computational flavor, at
least insofar as the entire realm of formal mathematics can be
viewed as being a fundamentally ``computational'' entity.  As one
interesting logical conclusion of this conceptual progression, if
all observed phenomena are indeed eventually explicable as being
aspects of some underlying purely mathematical/computational system,
then we can argue that in the end, there really is no need for a
separate \emph{physical} ontology at all any more; we could instead
validly suppose that the entire ``physical'' world really {\em is}
nothing but a certain (very elaborate and complex) abstract
mathematical or computational object.  Such a viewpoint has many
attractive philosophical features, at least from the perspective of
a hard-core rationalist.  One prominent proponent of such musings is
Tegmark, \eg, see \cite{Tegmark-98}.  Another proposal for unifying
mathematics and physics was recently made by Benioff
\cite{Benioff-02}.

However, regardless of one's personal feelings about such
far-ranging philosophical agendas, if we can at least show that it
is consistent to say that a given physical quantity can be exactly
identified with a given mathematical or computational quantity,
then, as scientists, we can certainly all agree that the most
parsimonious description of physics will indeed be one that does
make that identification, since otherwise our description of the
world would be burdened with an unnecessary proliferation of
artificially distinct concepts, in violation of Ockham's razor, the
most fundamental principle of scientific thought.

In this paper, we will primarily concern ourselves with just one
small aspect of the grander theme of interpreting physics as
information processing. Specifically, we focus on the idea of
interpreting the physical energy content of a given system as being
simply a measure of the rate at which that system is undergoing a
certain ubiquitous physical process---namely, quantum state
evolution---which can also be viewed as a computational process, as
we do in quantum computing.  In other words, the premise is that
physical energy is nothing but the \emph{rate of quantum computing},
if the meaning of this phrase is appropriately defined. This paper
will clarify precisely in what sense this statement is true.

We'll also see that the concept of physical \emph{action}, in a
certain (somewhat generalized) sense, corresponds to a computational
concept of the \emph{amount of computational effort exerted}, which
we'll call \defn{effort} for short.

Of course, it is not necessarily the case that a given system will
have been prepared in such a way that all of its physical
computational activity will actually be directly applied towards the
execution of a target application algorithm of interest.  In most
systems, only a small fraction of the system's energy will be
engaged in carrying out application logic on computational degrees
of freedom, while the rest will be devoted to various auxiliary
supporting purposes, such as maintaining the stability of the
machine's structure, dissipating excess heat to the environment,
\etc, or it may simply be wasted in some purposeless activity.

For that part of energy that \emph{is} directly engaged in carrying
out desired logical operations, we will see that one fruitful
application of the computational interpretation of energy will be in
allowing us to characterize the \emph{minimum} energy that must be
harnessed in order to carry out a given computational operation in a
given period of time.  In section \ref{sec:specops}, we will show
how to calculate this ``difficulty'' figure for a variety of simple
quantum logic operations, and we briefly discuss how to generalize
it to apply to classical reversible and irreversible Boolean
operations as well.

\section{Background}
Of course, the earliest hints about the relationship between energy
and the rate of computing can be found in Planck's original $E=h\nu$
relation for light, which tells us that an electromagnetic field
oscillation having a frequency of $\nu$ requires an energy at least
$h\nu$, where $h\simeq6.626\times10^{-34}\mathrm{J\,s}$ is Planck's
constant.  Alternatively, a unit of energy $E$, when devoted to a
single photonic quantum, results in an oscillation (which can be
considered to be a very simple kind of computational process)
occurring at a cycle rate of $\nu=E/h$.

Also suggestive is the Heisenberg energy-time uncertainty principle
$\Delta E\Delta t\geq h/2$, which relates the standard deviation or
uncertainty in energy $\Delta E$ to the minimum time interval
$\Delta t$ required to measure energy with that precision; the
measurement process can be considered a type of computation.
However, this relation by itself only suggests that the {\em spread}
or standard deviation of energy has something to do with the rate of
a process of interest; whereas we are also interested in finding a
computational meaning for the absolute or mean value of the energy,
itself.

More recently, in 1992, Tyagi {\cite{Tyagi-92}} proposed a notion of
``computational action'' that was based on the amount of energy
{\emph{dissipated}} multiplied by the elapsed time (a quantity which
has the same physical units as action) and proposed a theory of
optimal algorithm design based on a ``principle of least
computational action.'' However, Tyagi's analogy with Hamilton's
principle was still a long way from indicating that
{\emph{physical}} action actually {\emph{is}} computation in some
sense, or that physical energy itself (which is, in general, not
necessarily dissipated) corresponds to a rate of computation.
Still, it was suggestive.

Going much further, in 1998 Toffoli {\cite{Toffoli-98}} argued that
the least-action principle in physics itself can be derived
mathematically from {\emph{first principles}} (rather than as an
{\foreign{ad hoc}} physical postulate) as a simple combinatorial
consequence of counting the number of possible fine-grained discrete
dynamical laws that are consistent with a given macroscopic
trajectory.  In Toffoli's model, which intriguingly even captures
aspects of relativistic behavior, the energy of a state is
conjectured to represent the logarithm of the length of its
dynamical orbit. Toffoli also gives a correspondence between
physical action and amount of computation that is more explicit than
Tyagi's, and in which the path with the least Lagrangian action is
the one with the greatest amount of ``unused'' or ``wasted''
computational capacity. In later papers following up on the present
one, we will show that indeed, Lagrangian action corresponds
negatively to the portion of the computational effort that does not
contribute to an object's active motion.

At around the same time as Toffoli's work, Margolus and Levitin
\cite{Margolus-Levitin-98} showed that in any quantum system, a
state with a quantum-average energy $E$ above the ground state of
the system takes at least time $\Delta t \geq t^- = h/4E$ to evolve
to an orthogonal state, along with a tighter bound of $\Delta t \geq
t^-_N = (N-1)h/2NE$ that is applicable to a trajectory that passes
through a cycle of $N$ mutually orthogonal states before returning
to the initial state. In the limit as $N\rightarrow\infty$,
$t^-_N\rightarrow h/2E$, twice the minimum time of $t^-=t^-_2$ which
applies to a cycle between 2 states. Both bounds are achievable in
principle, in freely constructed quantum systems.

In a widely-publicized paper in {\em Nature} in 2000, Lloyd
\cite{Lloyd-00} used the Margolus-Levitin result to calculate the
maximum performance of a 1 kg ``ultimate laptop,'' in a hypothetical
limiting scenario in which all of the machine's rest mass-energy is
devoted to carrying out a desired computation.

Two years later, Levitin, Toffoli and Walton \cite{Levitin+02}
investigated the minimum time to perform a specific quantum logic
operation, namely a CNOT (controlled-NOT) together with an arbitrary
phase rotation, in systems of a given energy $E$.

In 2003, Giovannetti, Lloyd and Maccone
\cite{Giovannetti+03,Giovannetti+03b} explored tighter limits on the
time required to reduce the fidelity between initial and final
states to a given level, taking into account the magnitudes of both
$E$ and $\Delta E$, the system's degree of entanglement, and the
number of interaction terms in the system's Hamiltonian.

Results such as the above suggest that energy might fruitfully be
{\em exactly} identified with the rate of raw, low-level
quantum-physical ``computing'' that is taking place within a given
physical system, in some appropriate sense, if only the quantity
``amount of computing'' could be defined accordingly.  We would like
to show that some well-defined and well-justified measure of the
rate at which ``computational effort'' (not necessarily useful) is
being exerted within any quantum system is indeed {\em exactly}
equal to the energy of that system.

\section{Preview}

In subsequent sections of this paper, we address the aforementioned
goal by proposing a well-defined, real-valued measure of the total
{\em amount of change} undergone over the course of {\em any}
continuous trajectory of a normalized state vector along the unit
sphere in Hilbert space.  This measure is simply given by the line
integral of the magnitude of the imaginary component of the inner
product between infinitesimally adjacent normalized state vectors
along the given path.  This quantity is invariant under any
time-independent change of basis, since the inner product itself is.
As we will show, it is also numerically equal to twice the
complex-plane area (relative to the origin) that is circumscribed or
``swept out'' by the coefficients of the basis vector components, in
any basis. For closed paths, this quantity is even invariant under
not only rotations but also translations of the complex plane.
Finally, our quantity can be perhaps most simply characterized as
being {\defn{the action of the Hamiltonian}} along the path; this is
to be contrasted with the usual action (of the Lagrangian), whose
precise computational meaning will be addressed in later work.

We propose that the above-described measure of ``amount of change''
is the most natural measure of the amount of computational
\emph{effort} exerted by a physical system as it undergoes a
specific trajectory.  For any pair of trajectory endpoints, the
effort has a well-defined minimum value over possible trajectories
which is obtained along a ``geodesic'' trajectory between the
endpoint states, thereby inducing a natural metric over the Hilbert
space.

We will show that in any quantum system, the instantaneous rate at
which change occurs (computational effort is exerted) for any state,
under any time-dependent Hamiltonian operator, is exactly given by
the (Hamiltonian) instantaneous average energy of the state. Thus,
the state's energy {\emph{is}} exactly its rate of computation, in
this sense.

We use the word ``effort'' here rather than ``work'' both (a) to
distinguish our concept from the usual technical meaning of work in
physics as being directed energy, and also (b) to connote that
effort is something that can be ineffectually wasted; \ie, it does
not necessarily correspond to \emph{useful} computational work
performed.  In fact, we will see that indefinitely large amounts of
effort could be expended (inefficiently) in carrying out any given
quantum computational task, \ie in accomplishing a given piece of
computational work.

Despite having no upper bound, our concept of effort turns out to
still be meaningful and useful for characterizing computational
tasks, since (as we will see) any given quantum or classical
computational operation does have a well-defined and non-trivial
{\emph{minimum}} required effort for worst-case inputs, which we
will call the {\emph{difficulty}} of the operation.  As we will see,
for any pair of unitaries $U_1, U_2$, the difficulty of the
operation $U_2U_1^{\dagger}$ that takes us from $U_1$ to $U_2$ gives
a natural distance metric over $\mathrm{U}_n$, the Lie group of
rank-$n$ unitary operators.

The difficulty of a computational operation, according to our
definitions, determines the minimum time required to perform it on
worst-case inputs of given energy, or (equivalently) the minimum
worst-case energy that must be devoted to a system in order to
perform the operation within a given time.  The difficulty thus
directly characterizes the computational complexity or ``cost'' of a
given operation, in the same ``energy-delay product'' units that are
popular in electrical engineering, but where the energy here refers
to the average instantaneous energy that is {\em invested} in
carrying out the computation, rather than to the amount of energy
that is {\em dissipated}.

\section{A Simple Example}
In this section, we start by presenting a simple, concrete example
in order to help motivate our later, more general definitions.
Consider any quantum system subject to a constant (time-independent)
Hamiltonian operator $H$.  Let $|\mathrm{G}\rangle$ and
$|\mathrm{E}\rangle$ be any normalized, non-degenerate pair of the
system's energy eigenstates.  The labels G and E here are meant to
suggest the ground and excited states of a non-degenerate two-state
system, but actually it is not necessary for purposes of this
example that there be no additional states of higher, lower, or
equal energy.

Since the Hamiltonian is only physically meaningful up to an
additive constant, let us adjust the eigenvalue corresponding to
vector $|\mathrm{G}\rangle$ to have value 0 (\ie\ let
$H|\mathrm{G}\rangle=0$), and then let $E$ denote the eigenvalue of
$|\mathrm{E}\rangle$ (\ie,
$H|\mathrm{E}\rangle=E|\mathrm{E}\rangle$). For example, for a
two-state system, we could let $H=(1+\sigma_z)E/2$ with the usual
definition of the Pauli $z$-axis spin operator $\sigma_z=[%
\begin{array}{rr}
  \scriptstyle 1 & \scriptstyle 0 \\[-6pt]
  \scriptstyle 0 & \scriptstyle -1
\end{array}%
]$; and let $|\mathrm{G}\rangle=[%
\begin{array}{c}
  \scriptstyle 0\\[-6pt]
  \scriptstyle 1
\end{array}%
]$ and $|\mathrm{E}\rangle=[%
\begin{array}{c}
  \scriptstyle 1\\[-6pt]
  \scriptstyle 0
\end{array}%
]$, thus we have that $H=|\mathrm{E}\rangle\langle\mathrm{E}|$ and
so $E=1$.

Now, consider the initial state $|\psi_0\rangle =
(|\mathrm{G}\rangle+|\mathrm{E}\rangle)/\sqrt{2}$ at time $t=0$, and
let it evolve over time under the influence of the system's
Hamiltonian, with $|\psi(t)\rangle = \e^{\imag
Ht/\hbar}|\psi_0\rangle$ denoting the state vector at time
$t$.\footnote{For convenience, we use the opposite of the ordinary
sign convention in the time-evolution operator.} Let $c_{|{\rm
G}\rangle}(t)$ and $c_{|\mathrm{E}\rangle}(t)$ denote
$\langle\mathrm{G}|\psi(t)\rangle$ and
$\langle\mathrm{E}|\psi(t)\rangle$ respectively, \ie, the components
(complex coefficients) of the state vector $|\psi(t)\rangle$ when
decomposed in an orthonormal basis that includes
${|\mathrm{G}\rangle,|\mathrm{E}\rangle}$ as basis vectors.

Initially, $c_{|\mathrm{G}\rangle}(t) = c_{|\mathrm{E}\rangle}(t) =
1/\sqrt{2}$. Over time, $c_{|\mathrm{E}\rangle}$ phase-rotates in
the complex plane in a circle about the origin, at an angular
velocity of $\omega_{|\mathrm{E}\rangle}=E/\hbar$.  In time $t =
2E/h$, it rotates by a total angle of $\theta=\pi$.  The area swept
out by the line between $c_{|\mathrm{E}\rangle}(t)$ and the origin
is $a_{|\mathrm{E}\rangle} =
\frac{1}{2}\pi|c_{|\mathrm{E}\rangle}|^2 = \pi/4$.  This is the area
of a semi-circular half-disc with radius $r_{|\mathrm{E}\rangle} =
|c_{|\mathrm{E}\rangle}| = 1/\sqrt{2}$. Meanwhile,
$c_{|\mathrm{G}\rangle}(t)$ is stationary and sweeps out zero area.
The total area swept out by both components is thus $a=\pi/4$.  This
evolution is depicted in figure 1.

\begin{figure}
\centerline{\psfig{file=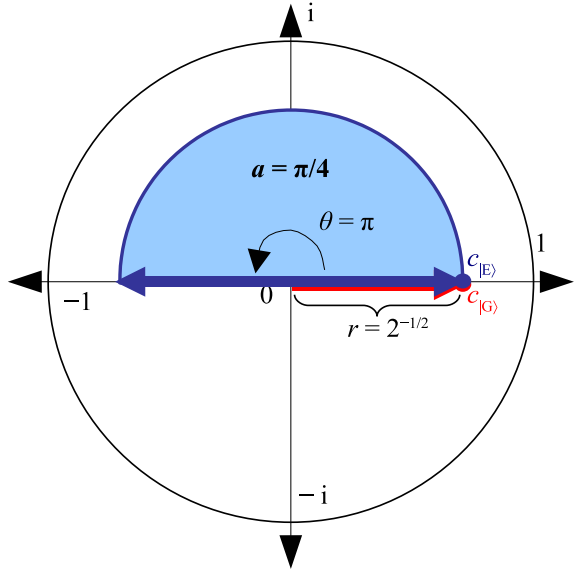,width=3in}}%
\caption{Under the Hamiltonian
$H=E|\mathrm{E}\rangle\langle\mathrm{E}|$, starting from the initial
state $\ket{\psi_0}=(\ket{\mathrm{G}}+\ket{\mathrm{E}})\cdot
2^{-1/2}$, the complex coefficient $c_{\ket{\mathrm{E}}} =
\langle\mathrm{E}|\psi(t)\rangle$ of $\ket{\mathrm{E}}$ (the excited
state) in the superposition sweeps out a half-circle in the complex
plane with area $\pi/4$ in time $t=2E/h$, while the ground-state
coefficient $c_{\ket{\mathrm{G}}}$ remains stationary.}
\label{fig:1}
\end{figure}

Does the area swept out by the complex components of the state
vector depend on the choice of basis?  We will answer this question
in a much more general setting later, but for now, consider, for
example, a new basis that includes basis vectors
$|\mathtt{0}\rangle$, $|\mathtt{1}\rangle$ where $|\mathtt{0}\rangle
= (|\mathrm{G}\rangle+|\mathrm{E}\rangle)/\sqrt{2}$ and
$|\mathtt{1}\rangle =
(|\mathrm{G}\rangle-|\mathrm{E}\rangle)/\sqrt{2}$. Consider the
evolution again starting from the same initial state as before,
$|\psi_0\rangle = |\mathtt{0}\rangle$. Note that the final state
after time $t=2E/h$ is $|\mathtt{1}\rangle$. In the new basis, the
coefficients $c_{|\mathtt{0}\rangle}(t)$ and
$c_{|\mathtt{1}\rangle}(t)$ respectively trace out the upper and
lower halves of a circle of radius $1/2$ centered at the point $1/2
+ \imag 0$.  The total area swept out by both components (on lines
between them and the origin) is the area of this circle, namely $a =
\pi(1/2)^2 = \pi/4$.  (See figure 2.)  Note that the total area in
this new basis is still $\pi/4$.

\begin{figure}
\centerline{\psfig{file=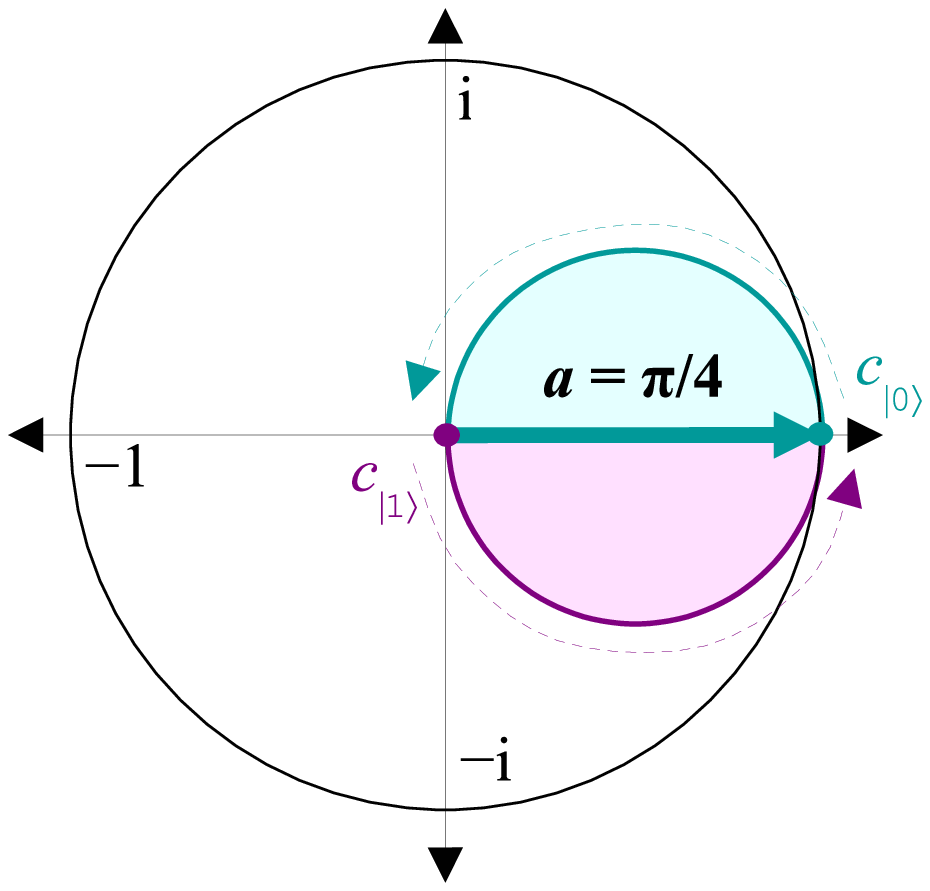,width=3in}}%
\caption{The evolution from figure 1, re-plotted in the basis
$\ket{\mathtt{0}}=(\ket{\mathrm{G}}+\ket{\mathrm{E}})\cdot
2^{-1/2}$,
$\ket{\mathtt{1}}=(\ket{\mathrm{G}}+\ket{\mathrm{E}})\cdot
2^{-1/2}$.  The coefficients of $\ket{\mathtt{0}}$ and
$\ket{\mathtt{1}}$ together sweep out a full circle, but the total
area swept out is still $\pi/4$.} \label{fig:2}
\end{figure}

At this point we may naturally ask, is the area the same in
{\emph{any}} fixed basis?  Later we will show that the answer is
yes; in general, the area swept out is independent of the basis for
{\em any} trajectory of {\em any} initial state.  The area swept out
will be (proportional to) our proposed measure of the amount of
computational effort exerted by a system in undergoing any specific
state-vector trajectory.

\section{General Framework}
In this section we proceed to set forth the general mathematical
definitions and notations to be used in the subsequent analysis.

\subsection{Time-independent case}
Let $\mathcal{H}$ be any Hilbert space.  Any linear,
norm-conserving, invertible, continuous and time-independent
dynamics on such a space must proceed via the application of a
unitary time-evolution operator, expressible as
\begin{equation}\label{eq:time-evolution}
    U = U(\Delta t) = \e^{\imag A(\Delta t)} = \e^{\imag H\Delta t}
\end{equation}
where $\Delta t$ is the length of a given time interval, $A(\Delta
t) = H\Delta t$ maps the interval to an Hermitian operator $A$ that
is proportional to $\Delta t$, and $H$ is an Hermitian operator with
units of angular frequency.  For any two times
$t_1,t_2\in\mathbb{R}$, and for any initial state vector $\ket{\psi}
= \ket{\psi(t_1)}$ at time $t_1$, the implied state at any other
time $t_2$ is given by $\ket{\psi(t_2)} = U(\Delta
t)\ket{\psi(t_1)}$, where $\Delta t = t_2-t_1$.  We will sometimes
also write $U$ and $A$ as functions of the directed pair of times,
written $t_1\rightarrow t_2$.  We will sometimes call the $U$ and
$A$ operators ``cumulative'' when the interval $\Delta t$ is not
infinitesimal.

Note that in eq.~(\ref{eq:time-evolution}) we are using the opposite
of the usual (but arbitrary) negative-sign convention in the
exponent; this is an inessential but convenient choice, in that
later it will let us automatically associate positive energies with
positive (\ie, counter-clockwise) phase velocities for the
coefficients of state components.

For convenience, for any operator $O$ and vector $v$, we will
sometimes use the notation $O[v]$ as an abbreviation for the
expectation value $\langle v|O|v\rangle$.

Now, of course, the eigenvectors of $U$ are also eigenvectors of $A$
and $H$, so $H$'s expectation value $H[\psi]$ for any initial vector
$\psi(t_1)\in\mathcal{H}$ is preserved by the time-evolution
$\psi(t_1)\rightarrow \psi(t_2)$. This conserved quantity (whose
existence follows from time-independence even more generally, via
N\"{o}ther's theorem) is called the {\defn{Hamiltonian energy}} of
the system.  Although in our expressions it has the dimensions of
angular velocity, this is the same as energy if we choose units
where $\hbar=1$, as is customary. Thus, $H$ is called the
Hamiltonian operator.  We will call the operator $A=A(t_1\rightarrow
t_2)$ the
\defn{cumulative action of the Hamiltonian from time $t_1$ to $t_2$},
where some of the qualifying phrases may be omitted for brevity. The
reasons for the use of the word ``action'' will be discussed later.

For convenience in the subsequent discussions, we will often just
set $t_1=0$ (without loss of generality) and write
$U=U(t)=U(0\rightarrow t) = \e^{\imag Ht}$. We refer to the complete
operator-valued function $\lambda t.U(t)$ for all $t$ values in some
range (which usually includes $t=0$, for which $U(0)=I$) as a
\emph{unitary trajectory} over that time interval.  Also, for any
$t$ we write $A(t):\equiv A(0\rightarrow t)$ for the cumulative
action from 0 to $t$.

Differentiating $U(t)$ with respect to time and applying the result
to an initial state $\ket{\psi(0)}$ then yields us Schr\"{o}dinger's
equation in various forms that we'll use,
\begin{eqnarray}
\dot{U} = \frac{\diff U(t)}{\diff t} = \frac{\diff}{\diff t}\e^{\imag Ht} &=& \imag H\e^{\imag Ht} = \imag HU(t) \label{eq:diffU} \\
\frac{\diff}{\diff t}U(t)|\psi(0)\rangle &=& \imag HU(t)|\psi(0)\rangle \\
\dot{|\psi\rangle} = \frac{\diff}{\diff t}|\psi(t)\rangle &=& \imag H|\psi(t)\rangle \label{eq:schrod} \\
    \frac{\diff}{\diff t} &=& \imag H,\label{eq:schrodop}
\end{eqnarray}
where again, note that we are using $\hbar=1$ and the opposite of
the usual sign convention.  Note also that we are able to
differentiate $\e^{\imag Ht}$ in eq.~(\ref{eq:diffU}) because
$\diff/\diff t$ commutes with $H$, since $H$ here is a constant.

\subsection{Time-dependent case}
The natural generalization of eq.~(\ref{eq:schrodop}) (the operator
form of Schr\"{o}dinger's equation) to a system with a
time-dependent Hamiltonian $H(t)$ is of course just
\begin{equation}\label{eq:timedep}
    \frac{\diff}{\diff t} = \imag H(t)
\end{equation}
where now $H(t)$ is permitted to vary over time, though often with a
constraint that it be differentiable, smooth, or analytic.

One may at first think that in this time-dependent context, we could
appropriately generalize the time-evolution operator equation
(\ref{eq:time-evolution}) by simply changing the definition of the
action operator $A$ (as a function of $t$) from the original
$A(t)=Ht$ to what one might na\"{i}vely think would be the obvious
generalization to a time-dependent $H$,
\begin{equation}\label{eq:bad-A}
    A(t) = \int_{\tau=0}^t H(\tau) \diff \tau,
\end{equation}
while still keeping the relation $U(t) = \e^{\imag A(t)}$.  But in
fact, the definition (\ref{eq:bad-A}) does not work for this
purpose, since in general the values of $H(\tau)$ at different times
$\tau$ will not commute with each other; taking the integral loses
all information about their relative time-ordering, and the
time-derivative of $U(t)$ will no longer be equal to $\imag H(t)$ as
required, since $\diff/\diff t$ will no longer commute with $H(t)$.

The standard way to repair this problem (discussed in almost any
quantum field theory textbook, \eg, \cite{Sterman-93}) is to define
a time-ordering meta-operator $\mathcal{T}$, which takes a given
operator expression and reorders its internal operator products so
that operators associated with earlier time points are applied first
in all products (reading right-to-left).  For example, as a matter
of definition,
\begin{equation}
    \mathcal{T}[H(t_1)H(t_2)] :\equiv \left\{\begin{array}{ll}
                                H(t_1)H(t_2) & \mbox{if $t_1>t_2$} \\
                                H(t_2)H(t_1) & \mbox{otherwise}
                            \end{array}
                            \right. \label{eq:time-order}
\end{equation}
With this notational convention, we can write
\begin{equation}
    U(t) = \mathcal{T}\e^{\imag A(t)} \label{eq:time-ord-exp}
\end{equation}
where $A(t)$ is as defined in eq.~(\ref{eq:bad-A}), and the meaning
of this meta-expression will be well-defined and consistent with
eq.~(\ref{eq:timedep}) applied to $U(t)$.  But the problem with this
approach is that the expression $A(t)$ in (\ref{eq:time-ord-exp}) no
longer denotes a ``first class object'' of our language, but rather
is a sort of meta-mathematical place-holder to be manipulated via a
rather complex interpretational procedure, which involves applying
eq.~(\ref{eq:time-order}) to uncountably many infinitesimal pieces
of the integrals appearing in the Taylor-expanded version of
eq.~(\ref{eq:time-ord-exp}).  There is no longer any simple, direct
relationship between the properties of the linear operator $A(t)$
defined in eq.~(\ref{eq:bad-A}) (\eg, its eigenvalues and
eigenvectors) and the properties of $U(t)$.

Thus, in what follows we will find it more useful to instead abandon
eq.~(\ref{eq:bad-A}), and take the rather more concrete approach of
simply redefining $A(t)$ for a given unitary trajectory $U(t)$ to be
the unique continuously time-dependent Hermitian operator such that
$A(0)=0$ and
\begin{equation}
    U(t) = \e^{\imag A(t)} \label{eq:action}
\end{equation}
(with \emph{no} time-ordering operator!) for all $t$.  To see that
such an $A$ indeed exists and is unique, note that since each
particular $U = U(t)$ (at a given moment) is unitary, it is a normal
operator and can thus be given a spectral decomposition
\begin{equation}
    U = \sum_i u_i|u_i\rangle\langle u_i|
\end{equation}
where $\{|u_i\rangle\}$ and $\{u_i\}$ respectively comprise an
orthonormal eigenbasis of $U$ and the corresponding unit-modulus
eigenvalues. We can therefore define the multi-valued logarithm of
$U$ by
\begin{eqnarray}
    \ln U &=& \ln \sum_i u_i|u_i\rangle\langle u_i| \nonumber \\
          &:\equiv& \sum_i (\ln u_i)|u_i\rangle\langle u_i| \nonumber \\
          &=& \sum_i \imag\arg(u_i)|u_i\rangle\langle u_i| \label{eq:logU3} \\
          &=& \sum_i \imag[\mathrm{Arg}(u_i) + 2\pi n_i]|u_i\rangle\langle u_i|
          \label{eq:logU}
\end{eqnarray}
where in step~(\ref{eq:logU3}) we have used the fact that $|u_i|=1$,
and where in line~(\ref{eq:logU}) $\mathrm{Arg}(u_i)\in[0,2\pi)$
denotes the principal value of the multivalued function $\arg(u_i)$,
while the $n_i$ values may be any integers. Although we see that
there are infinitely many values of $(\ln U)$ for any individual $U$
in isolation, nevertheless there \emph{is} a unique single-valued
definition of the entire function $L(t) = \ln U(t)$, given the
function $U(t)$, that is \emph{continuous} over $t$ and where
$L(0)=0$.

The uniqueness is due to the fact that $U(t)$ varies continuously in
$t$, and thus, if we like, the eigenbasis $\{|u_i(t)\rangle\}$ that
we choose for $U$ at each moment (which has $k$ free gauge-like
parameters determining the $u_i$, where $k=\dim\mathcal{H}$) can
vary continuously as well. Given basis vectors $\ket{u_i}$ (and thus
$u_i$ values) that change continuously, it follows that at any
moment, only one assignment of values to the ${n_i}$ parameters can
possibly yield continuity with the logarithm value $L(t-\diff t)$ at
the previous moment, since any other choice would (discontinuously)
change one of the phase angles $\mathrm{Arg}(u_i) + 2\pi n_i$ in the
expression (\ref{eq:logU}) by an amount that is (infinitesimally
close to) a multiple of $2\pi$.  The $n_i$ parameters can (and must)
change by $\pm 1$ from their preceding values (while leaving $L(t)$
continuous) only at a discrete set of time points, namely those
where the continuously-changing $u_i$ value crosses the branch cut
of the Arg() function (in some direction), and $\mathrm{Arg}(u_i)$
jumps by $\mp 2\pi$.

Now, given this uniquely-defined unitary trajectory logarithm $L(t)
= \ln U(t)$, we simply define our action operator as $A(t) = -\imag
L(t)$, and then trivially we have that $U(t) = \e^{\imag A(t)}$
holds for all $t$, where the exponential can be defined via the
spectral decomposition of $A$ (equivalently to the standard
Taylor-series definition), thereby inverting the logarithm.

Meanwhile, the entire unitary trajectory $U(t)$ itself is derived
from the Hamiltonian trajectory $H(t)$ by setting $U(0)=I$ and
applying the operator form (\ref{eq:timedep}) of the time-dependent
Schr\"{o}dinger equation to $U(t)$.  So $(\diff/\diff t)U(t) = \imag
H(t)U(t)$, and we are thereby guaranteed that in fact
\begin{equation}
    \frac{\diff}{\diff t}\e^{\imag A(t)} = \imag H(t)\e^{\imag A(t)}
\end{equation}
as desired, which (recall) failed to be true (in the absence of a
time-ordering operator) for the $A(t)$ defined in
eq.~(\ref{eq:bad-A}).

For reasons we will explain, we will refer to a complete function
$\lambda t.A(t)$ as defined by eq.~(\ref{eq:action}) as the
{\defn{cumulative Hamiltonian action trajectory}} implied by the
Hamiltonian trajectory $H(t)$.

In cases where $H(t) = H$ is constant over time, note that this
definition of $A(t)$ reduces to the simple $Ht$ form that we used
back in eq.~(\ref{eq:time-evolution}).  This follows from the
observation that the definition $A(t)=Ht$ indeed solves
eq.~(\ref{eq:action}) when $H$ is constant, and the fact that (as we
just showed) the $A(t)$ implied by eq.~(\ref{eq:action}) is unique
under the continuity constraint.

Later, we will see the importance of the Hamiltonian action
trajectory $A(t)$, and discuss the precise meaning and computational
interpretation of its expectation value when applied to a given
state.

To clarify our terminology, note that in this document we are using
the word \emph{action} in a somewhat more general sense than is
usual; typically in physics (\eg, in Hamilton's principle)
``action'' just refers to the quantity having units of action that
is obtained by integrating the Lagrangian $L=pv-H$ along some path.
However, it is also perfectly valid and reasonable to consider the
more general notion of the action that is associated with
{\emph{any}} quantity that has units of energy, by setting the
time-derivative of that action along some path to be equal to that
energy.

Indeed, we will see later that the time-derivative of the cumulative
Hamiltonian action $A(t)$ (as we have defined it) along a given
trajectory is in fact exactly the instantaneous Hamiltonian energy
$H(t)$, \ie,
\begin{equation}
\frac{\diff}{\diff t}A(t)[\psi(0)] = H(t)[\psi(t)],
\end{equation}
similarly to how the time-derivative of the ordinary (\ie,
Lagrangian) action along a given trajectory is the instantaneous
Lagrangian energy $L(t)$.

As a final piece of notation which will help us generalize our
results to the time-dependent case, we will sometimes write $U'(t)$
to refer to the ``instantaneous'' unitary transformation that
applies over an infinitesimal time interval $\diff t$ at time $t$,
that is,
\begin{eqnarray}
U'(t) &:\equiv& U(t\rightarrow t+\diff t) \nonumber \\
      &=& 1 + \imag H(t)\diff t. \label{eq:instU}
\end{eqnarray}
Note also that any larger transformation $U(t_1\rightarrow t_2)$ can
be expressed as the time-ordered product of all the infinitesimal
$U'(t)$ over the continuum of times $t$ in the range from $t_1$ to
$t_2$.  That is, we can write
\begin{equation}\label{eq:contprod}
    U(t_1\rightarrow t_2) = \mathcal{T}\prod_{t=t_1}^{t_2}U'(t)
\end{equation}
with the opposite ordering if $t_2<t_1$. Thus, $U'(t)$ uniquely
defines $U(t)$, so we will sometimes refer to $U'(t)$ as the unitary
trajectory also.

We should keep in mind that although the complete unitary trajectory
$U(t)$ (or $U'(t)$) between $t_1$ and $t_2$ determines the overall
transformation $U(t_1\rightarrow t_2)$, the converse is not true:
Knowing the cumulative $U=U(t_1\rightarrow t_2)$ for a particular
pair of times $t_1,t_2$ is of course insufficient to determine a
unique unitary trajectory $U(t)$, since in general infinitely many
cumulative action operators $A=A(t_1\rightarrow t_2)$ can
exponentiate to yield the same cumulative $U$ (since
expression~(\ref{eq:logU}) is multivalued), and furthermore, in the
time-dependent case, a continuum of different Hamiltonian
trajectories $H(t)$ (which determine $U'(t)$) could implement a
given cumulative action operator $A$.

We will similarly use the notation $A'(t) = H(t)\diff t$ to denote
the infinitesimal action operator that applies from time $t$ to
$t+\diff t$; note that $U'(t) = \e^{\imag A'(t)} = 1 + \imag
H(t)\diff t$.

\section{Defining Computational Effort}
With the above general definitions and observations aside, let us
now proceed to define our concept of the amount of computational
effort exerted by a system in undergoing a state trajectory
$\ket{\psi(t)}$ between two times.

We will find it easiest to define this quantity first for the case
of a system with a time-independent Hamiltonian
$H(t)=H=\mathrm{const}$. Later, we will show how our results can be
generalized to the time-dependent case.

Let $\ket{v}$ be any eigenvector of $H$, and $\omega$ the
corresponding eigenvalue, which is real since $H$ is Hermitian. That
is, let $H\ket{v} = \omega\ket{v}$. Thus, $\ket{v}$ is also an
eigenvector of the cumulative action operator $A(t)=Ht$ for any $t$,
with eigenvalue $\alpha=\omega t$.

First, when $t$ is an infinitesimal $\diff t$, consider the
instantaneous $U' = 1+\imag H\diff t$.  Clearly, $\ket{v}$ is an
eigenvector of $U'$, since $U'\ket{v} = (1+\imag H\diff t)\ket{v} =
(1+\imag \omega\diff t)\ket{v} = u\ket{v}$, where the scalar
$u=1+\imag\omega\diff t = \e^{\imag\omega\diff t} =
\e^{\imag\diff\alpha}$. Thus, under application of $U'$, the
eigenvector $\ket{v}$ transforms to $\ket{v'} :\equiv
\e^{\imag\omega\diff t}\ket{v} = \e^{\imag\diff\alpha}\ket{v}$, that
is, it phase-rotates in the complex plane at angular velocity
$\omega$ through an infinitesimal angle $\diff\alpha$.  Note also
that
\begin{eqnarray}
\Im\langle v|v'\rangle &=&
    \Im\langle v|(1+\imag\diff\alpha)|v\rangle =
    \Im(1+\imag\diff\alpha)\langle v|v\rangle
    \nonumber \\
    &=& \diff\alpha = \E{\omega\diff t}{v} =
        \langle v|A'|v\rangle = A'[v]. \label{eq:imagadj}
\end{eqnarray}
That is, when $\ket{v}$ is an eigenvector of $H$, the magnitude of
the imaginary part of the inner product between infinitesimally
adjacent state vectors is equal to the expectation value $A'[v]$ of
the infinitesimal action operator $A'=H\diff t$ applied to the
state. As we go on, we will extend the
relationship~(\ref{eq:imagadj}) to non-infinitesimal trajectories,
non-eigenvectors, and time-dependent Hamiltonians.

Next, note that the eigenvectors $\ket{v}$ of $H$ are also
eigenvectors of the cumulative action operators $A(t)=Ht$ and
cumulative unitaries $U(t) = \e^{\imag A(t)} = \e^{\imag Ht}$, and
vice-versa. Let $A(t)\ket{v}=\alpha(t)\ket{v}$, with $\ket{v}$ a
fixed eigenket of $A(t)$, and with $\alpha(t)=\omega t$ as its
eigenvalue. Then, $U(t)\ket{v} = \e^{\imag A(t)}\ket{v} = \e^{\imag
\alpha(t)}\ket{v} = u(t)\ket{v}$ where $u(t)=\e^{\imag \alpha(t)}$.
Thus, upon the application of $U$, $\ket{v}$ gets multiplied by the
phase factor $u(t)$, or (we can say) rotated by a total phase angle
of $\alpha(t) = \omega t$, which could be much greater than $2\pi$
in long evolutions, as can also be seen by integrating $\diff\alpha$
over $t$.  Note also that if we integrate $\Im\langle v|v'\rangle$
along the trajectory, we still get the cumulative action
$A(t)[v(0)]$:
\begin{eqnarray}
\int_{\tau=0}^{t} \Im\langle v(\tau)|v'(\tau)\rangle
    &=& \int_{\tau=0}^t \Im\langle
        v(\tau)|(1+\imag\omega\diff\tau)|v(\tau)\rangle \\
    &=& \omega t = \alpha(t) = \langle v(0)|A(t)|v(0)\rangle.
\end{eqnarray}

Next, consider an arbitrary pure state $\ket{\psi(0)} = \sum_i
c_i(0)\ket{v_i}$, where the $\ket{v_i}$ are normalized eigenstates
of $H$ with eigenvalues $\omega_i$, and the $c_i(0)$ are the initial
coefficients of the $\ket{v_i}$ in the superposition. The state at
time $t$ can be expressed as
\begin{eqnarray}
\ket{\psi(t)} &=& \sum_i \exp[\imag\alpha_i]c_i(0)\ket{v_i}     \nonumber \\
        &=& \sum_i \exp[\imag\omega_i t]c_i(0)\ket{v_i}  \nonumber \\
        &=& \sum_i c_i(t)\ket{v_i},
\end{eqnarray}
where we see that each coefficient $c_i(t) = \exp[\imag\omega_i
t]c_0(t)$ (in the fixed basis $\{\ket{v_i}\}$) simply phase-rotates
with angular velocity $\omega_i$ along an origin-centered circle in
the complex plane with constant radius $r_i=|c_i|$.  Over any amount
of time $t$, we see that $c_i$ rotates in the complex plane by a
total angle of $\alpha_i = \omega_i t$, while the line in the
complex plane that joins $c_i$ to the origin sweeps out an arc with
an area of $a_i=\frac{1}{2}\omega_i t r_i^2$. (See figure 3 for an
illustration of the area swept out in the infinitesimal case.)  For
example, in time $t = 2\pi/\omega_i$, coefficient $c_i$ sweeps out a
complete disc of area $a_i=\pi r_i^2$ as it traverses an angle of
$\alpha=2\pi$.  For consistency, in the case of clockwise rotations
(negative $\omega_i$), we will consider the area swept out to also
be negative.

\begin{figure}
\centerline{\psfig{file=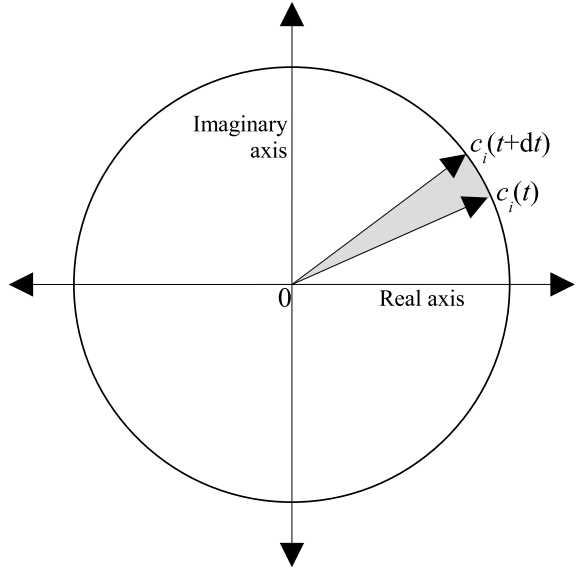,width=3in}}%
\caption{In the energy eigenbasis, a complex coefficient $c_i$ of a
basis state sweeps out a small wedge-shaped area (shown exaggerated)
in the complex plane over an infinitesimal time interval $\diff t$.}
\label{fig:3}
\end{figure}

Now, let $\psi'(t)=\psi(t+\diff t)$. Then
\begin{eqnarray}
\int_{\tau = 0}^{t} \Im\langle\psi(\tau)|\psi'(\tau)\rangle
    &=& \int_{\tau = 0}^{t} \Im\sum_i
        \bar{c}_i(\tau)c_i(\tau+\diff\tau) \\
    &=& \int \sum_i
        r_i^2 \Im\{
        \e^{-\imag\theta_i(\tau)}\e^{\imag[\theta_i(\tau)+\omega_i\diff\tau]}\} \\
    &=& \int \sum_i p_i\Im(1+\imag\omega_i\diff\tau) \\
    &=& \int \sum_i p_i\diff\alpha_i \\
    &=& \int \diff\alpha = \alpha(t) = A(t)[\psi(0)]
\end{eqnarray}
where the overbar denotes complex conjugation, $r_i = |c_i|$ as
before, $\theta_i(\tau) = \arg(c_i(\tau))$, and $\alpha$ is now the
weighted-average value of $\alpha_i$.

Now, consider the \emph{total} area $a(t)$ swept out by \emph{all}
coefficients $c_i$ over time $t$.  Note that $r_i^2=|c_i|^2$ is also
the probability $p_i$ of basis state $v_i$, and so the
{\emph{total}} area swept out is always exactly half of the
{\emph{average}} angle $\alpha(t)$ of phase rotation (weighted the
by state probability), or in other words, half of the expectation
value of the $A(t)$ operator applied to the state $\psi(0)$. That
is,
\begin{eqnarray}
a(t) &=& \sum_i \frac{1}{2}\omega_i t r_i^2               \nonumber \\
     &=& \frac{1}{2} \sum_i p_i \alpha_i                   \nonumber \\
     &=& \frac{1}{2} A(t)[\psi(0)] = \frac{1}{2} \alpha(t).
\end{eqnarray}

Thus we have shown that for time-independent Hamiltonians, the
expectation value of the action operator $A(t)$ applied to any
initial state $\psi(0)$ is equal to the integral over the state
trajectory of the inner product between infinitesimally adjacent
states $\psi(t)$ and $\psi'(t) = \psi(t+\diff t)$ along the
trajectory, as well as to the average phase angle $\alpha$
accumulated and to twice the complex-plane area $a$ swept out by the
state's coefficients, when the state is decomposed in the energy
eigenbasis.

Of course, the inner product between two state vectors is a pure
geometric quantity, and so is basis-independent.  Therefore, the
integral of $\Im\langle\psi|\psi'\rangle$ over the state trajectory
does not depend at all on the (fixed) choice of basis under which
states are decomposed into components.  Likewise, the operator
$A(t)$ itself is a geometric object not inherently associated with
any particular basis.  Therefore, the identity
\begin{equation}
\int_{\tau=0}^t \Im\langle\psi(\tau)|\psi'(\tau)\rangle =
 A(t)[\psi(0)]
\end{equation}
that we proved above is a fundamental one whose truth does not rely
on any particular basis or coordinate system.

However, it is perhaps somewhat less obvious that the average angle
$\alpha$ of phase rotation and the complex-plane area $a$ swept out
by the state coefficients should also be basis-independent
quantities, since their original definitions explicitly invoked a
choice of basis (the energy basis).  However, in the next section we
will show that in fact, these quantities are basis-independent as
well. Thus, all of the following identities still hold true,
regardless of basis:
\begin{equation}
2a = \alpha = \int_{\tau=0}^{t}\Im\langle \psi|\psi'\rangle =
A(t)[\psi(0)],
\end{equation}
where $a$ is the total complex-plane area swept out by the state
coefficients in any fixed basis, $\alpha = \int \omega\diff t$ is
the time-integral of the expected value $\omega$ of the angular
velocity $\omega_i$ of the state coefficients in any fixed basis
(not necessarily the same one), $\psi=\psi(\tau)$ is the state
trajectory, with $\psi'=\psi(\tau+\diff\tau)$, $A(t)$ is the action
operator as we defined in equation~(\ref{eq:action}), and we are
using our mean-value notation $A(t)[\psi(0)]=\langle
\psi(0)|A(t)|\psi(0)\rangle$.

Our proposed measure of the amount of change undergone (and
computational effort exerted) along a state trajectory $\psi(t)$
generated by a constant $H$ will then just be the $\alpha$ value for
that trajectory.

Later, in section 8, we will show that the above identities also
still hold even when $H(t)$ varies over time, and so our measure
will generalize to that case as well.

\section{Generalizing to Arbitrary Bases}
The above discussion made use of a set of basis vectors
$\{\ket{v_i}\}$ which were taken to be orthonormal eigenvectors of
the (temporarily presumed constant) Hamiltonian operator $H$.  Now,
we will show that this particular choice of basis was in fact
unnecessary, and that the same statements concerning the
relationship between the area swept out, the average phase angle
accumulated, and the action $A(t)$ would remain true in any fixed
(time-independent) basis.

At first, it may seem very non-obvious that the area swept out
should still be exactly half of the action.  Note that our previous
arguments for this relied on the fact that in the energy basis
$\{\ket{v_i}\}$, the coefficients $c_i$ all rotate at uniform
angular velocities $\omega_i$ in circles in the complex plane, while
their individual magnitudes remain constant.  In a different basis
${\ket{v_j}}$ (distinguished by using a different index symbol $j$),
this will no longer be true. Each basis vector $\ket{v_j}$ in the
new basis is in general some superposition of the $\{\ket{v_i}\}$,
such as
\begin{equation}\label{veejay}
\ket{v_j} = \sum_i u_j^i \ket{v_i},
\end{equation}
where the matrix $\mathbf{U}=[u_j^i]$ of complex coefficients (with
the subscript $j$ indexing rows, and the superscript $i$ indexing
columns) is, most generally, any unitary matrix. We can also write
this equation in matrix-vector form as
$\overrightarrow{\ket{v_j}}=\mathbf{U}\overrightarrow{\ket{v_i}}$,
where the over-arrow here denotes that we are referring to the
entire
column-ordered sequence of basis vectors, $\overrightarrow{\ket{v_i}}=\left[%
\begin{array}{c}
 \ket{v_1} \\[-6pt]
 \vdots \\
\end{array}\right]$.
Of course, a general state vector $\psi$ can equally well be
expressed as a linear superposition of either set of basis vectors,
that is,
\begin{eqnarray}
\ket{\psi} &=& \sum_ic_i\ket{v_i} \label{viexpand}\\
\ket{\psi} &=& \sum_jc_j\ket{v_j}. \label{vjexpand}
\end{eqnarray}
But now, we can substitute eq.~(\ref{veejay}) into
eq.~(\ref{vjexpand}) and rearrange, as follows:
\begin{equation}\label{viexpand2}
\ket{\psi} = \sum_{ij}c_ju_j^i\ket{v_i} = \sum_i\left(\sum_j
c_ju_j^i\right)\ket{v_i}.
\end{equation}
Now, since the ${\ket{v_i}}$ are linearly independent, the expansion
of $\ket{\psi}$ in terms of them must be unique, so we can equate
the coefficients on $\ket{v_i}$ in equations (\ref{viexpand}) and
(\ref{viexpand2}) to get
\begin{eqnarray}
c_i &=& \sum_i u_j^i c_j \nonumber \\
\overrightarrow{c_i} &=& \mathbf{U}^\mathrm{T}\overrightarrow{c_j},
\end{eqnarray}
where T is matrix transpose.  We can easily solve this equation for
the $c_j$ coefficients as follows:
\begin{eqnarray}
\overrightarrow{c_i} &=& \mathbf{U}^\mathrm{T}\overrightarrow{c_j} \nonumber \\
(\mathbf{U}^\mathrm{T})^{-1}\overrightarrow{c_i} &=& \overrightarrow{c_j} \nonumber \\
\bar{\mathbf{U}}\overrightarrow{c_i} &=& \overrightarrow{c_j} \nonumber \\
c_j &=& \sum_i \bar{u}_j^ic_i.
\end{eqnarray}
In other words, each complex coefficient in the new basis is just a
particular linear combination of what the various complex
coefficients were in the old basis.

If the coefficients $c_i$ in the old energy basis are describing
perfect circles around the complex origin at a variety of radii and
angular velocities, there is no guarantee that the coefficients
$c_j$ in the new basis will still be describing circular paths
centered on the origin, although their paths will of course still be
continuous and smooth if the original $c_i$ trajectories were.  In
general, the $c_j$ will follow complicated looping trajectories in
the complex plane, generated as if by Ptolemaic planetary epicycles,
\ie, as a sum of circularly rotating vectors.  A given $c_j$ will in
general return to its initial location in the complex plane only
when its components $c_i$ that have nonzero values of $u_j^i$ all
simultaneously return to their initial locations exactly, which
might even take infinitely long, if the corresponding $\omega_i$
values were relatively irrational.

Anyhow, the important point for our present purposes is that the
$c_j$s do not, in general, maintain a constant magnitude (distance
from the origin), and so the area swept out by the $c_j$ over a
given time is no longer just a section of a circle, which was very
easy to analyze.  Instead, while $c_j$'s phase angle $\theta_j$ is
rotating, simultaneously its magnitude $r_j$ may also be growing or
shrinking. Figure 4 illustrates the situation.

\begin{figure}
\centerline{\psfig{file=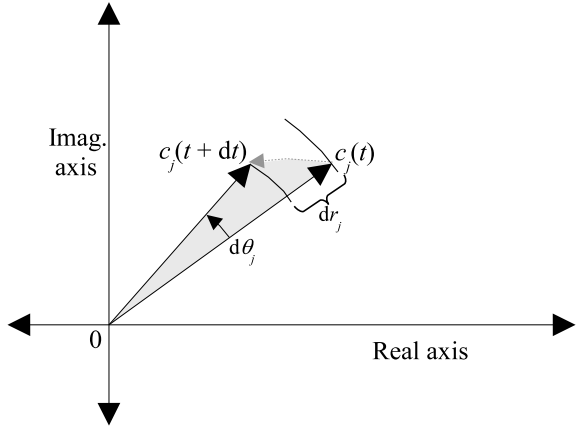,width=3in}}%
\caption{Area swept out (exaggerated) by a coefficient $c_j$ (in a
basis other than the energy eigenbasis) over an infinitesimal time
interval $\mathrm{d}t$.  Note that both its phase and its magnitude
change, in general.} \label{fig:4}
\end{figure}

To clarify what we mean by the phase angle $\theta_j(t)$ a bit more
carefully, let us use $\diff\alpha_j(t)\approx 0$ to denote the
infinitesimal increment of phase angle from times $t$ to $t+\diff t$
such that
\begin{equation}
\diff\alpha_j \equiv \arg(c_j')-\arg(c_j) \pmod{2\pi},
\end{equation}
so that $\diff\alpha_j$ remains infinitesimal even when $c_j$
crosses a branch cut of the Arg() function. Then, let $\alpha_j(t)$
be the total accumulated phase angle over time $t$, that is, the
integral of $\diff\alpha_j$ over time,
\begin{equation}
\alpha_j(t) = \int_{\tau=0}^t \diff\alpha_j
\end{equation}
so that $\alpha_j(0) = 0$.  Now, just let $\theta_j(t) =
\mathrm{Arg}[c_j(0)] + \alpha_j(t)$.  Thus also
$\diff\theta_j=\diff\alpha_j$.

What, now, is the area swept out in our new basis? First, notice
that in the infinitesimal limit, it is exactly half of the area of
the parallelogram that is spanned on two adjacent sides by $c_j =
c_j(t)$ and $c'_j = c_j(t+\diff t)$, considered as vectors in the
complex plane.  See figure 5.

\begin{figure}
\centerline{\psfig{file=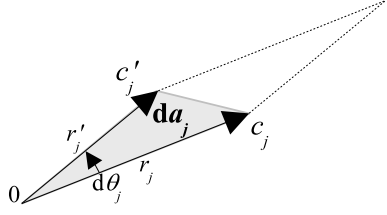,width=2in}}%
\caption{The infinitesimal area $\diff a_j$ swept out approaches
one-half of the parallelogram area $r_jr_j'\sin\diff\theta_j$.}
\label{fig:5}
\end{figure}

The parallelogram area, itself, is $\diff a_j = r_j
r'_j\sin(\diff\theta_j)$, where $r_j$ and $r'_j$ are the magnitudes
of the old and new coefficients, respectively.  However, note that
the area $\diff a_j$ of this parallelogram is also the signed
magnitude of the scalar ``cross product'' $c_j\times c'_j$ between
the coefficients, considered as vectors in the complex plane. (The
traditional cross product, defined in three dimensions, would be a
vector perpendicular to the complex plane having this value $\diff
a_j$ as its length.)  There is a nice identity \cite{Needham-97}
connecting the scalar cross product and dot product with the
conjugate multiplication of complex numbers, namely:
\begin{equation}
\bar{c}d = c\cdot d + \,\imag(c\times d),
\end{equation}
where $\bar{c}$ means the complex conjugate of $c$, and $c\cdot d$
denotes the real scalar ``dot product'' between $c$ and $d$
considered as vectors, namely $|c||d|\cos[\arg(d)-\arg(c)]$, and
$c\times d$ denotes the real scalar ``cross product'' previously
mentioned, namely $|c||d|\sin[\arg(d)-\arg(c)]$.

Applying this identity to our situation, we can see that the area
swept out, since it is half the cross product, is half of the
imaginary part of the conjugate product $\bar{c}_jc'_j$ between the
old and new coefficients, and also to half of
$\sin(\diff\alpha_j)=\diff\alpha_j$;
\begin{equation}
\diff a_j = \frac{1}{2}\diff\alpha_j =
\frac{1}{2}\Im(\bar{c_j}c'_j).
\end{equation}

Now, this is just the area swept out by a single component $c_j$. To
find the total area $\diff a$ swept out by all coefficients, we
merely sum over components:
\begin{eqnarray}
\diff a &=& \frac{1}{2}\sum_j \Im(\bar{c}_jc'_j)
   = \frac{1}{2}\Im\sum_j \bar{c}_jc'_j \nonumber \\
 &=& \frac{1}{2}\Im\langle \psi|\psi'\rangle =
 \frac{1}{2}\diff\alpha \label{eq:basis-indep}
\end{eqnarray}
In other words, \emph{just like in the energy basis}, in an
arbitrary basis, it is still true that the infinitesimal increment
$\diff a$ in the area swept out by the coefficients is exactly
one-half of $\Im\langle \psi|\psi'\rangle$, the imaginary component
of the inner product between infinitesimally adjacent vectors
$\psi=\psi(t)$ and $\psi'=\psi(t+\diff t)$ along the trajectory, and
further that this is equal to half of $\diff\alpha=\diff\theta$, the
average increment of the continuously-varying phase angles
$\theta_j(t)$ of the coefficients.

Now, we saw earlier that $\Im\langle \psi|\psi'\rangle$ is also
equal to the expectation value $A'[\psi]=\langle
\psi|A'|\psi\rangle$ of the infinitesimal action operator $A'=H\diff
t$ applied to the state $\psi$, for any state $\psi$.  So in
connection with the result (\ref{eq:basis-indep}) that we just
obtained, this means that $A'[\psi]$ gives {\emph{exactly}} the
average phase angle accumulation $\diff\alpha$ of the coefficients
$c_j$ of $\psi$ in {\emph{any}} basis, and twice the complex-plane
area $\diff a$ swept out by those coefficients. We can thus think of
$A'$ as being the operator representation of a fundamental,
basis-independent concept of ``average angle accumulated'' or
``total area swept out'' over infinitesimal intervals.

\section{Generalizing to Time-dependent Hamiltonians}
In the previous section, we established the basis-independence of
the identities $2\diff a = \diff\alpha = \Im\langle
\psi|\psi'\rangle=\omega\,\diff t = A'[\psi] = \langle \psi|H\diff t|\psi\rangle$
for infinitesimal changes of the state vector ($\psi\rightarrow \psi'$)
along its trajectory over infinitesimal time intervals $\diff t$,
under any \emph{constant} Hamiltonian $H$.

But, as long as the Hamiltonian $H(t)$ only changes in continuous
fashion, it can always be considered essentially ``constant''
throughout any infinitesimal interval $\diff t$, even if it is
varying over non-infinitesimal timescales. Therefore, the above
identities will still hold true instantaneously even for a
time-dependent Hamiltonian $H(t)$, which is what we originally
started out our discussion with. Thus, when we integrate the above
equation (\ref{eq:basis-indep}) over time, it remains true that:
\begin{eqnarray}
2a = \alpha &=& \int_{t=t_1}^{t_2}\Im\langle \psi(t)|\psi(t+\diff t)\rangle \label{eq:td1}\\
    &=& \int_{t=t_1}^{t_2}\omega(t)\diff t \label{eq:td2} \\
    &=& \int_{t=t_1}^{t_2} \langle \psi(t)|H(t)|\psi(t)\rangle \diff t\label{eq:td3} \\
    &=& \int_{t=t_1}^{t_2} A'(t)[\psi(t)].\label{eq:tdrel}
\end{eqnarray}
In words, this says that for any initial state $\psi$, we have that
$2a$ (twice the complex-plane area swept out by the coefficients of
$\psi$, in any basis) is equal to $\alpha$, the average phase angle
swept out by the state coefficients, as well as to (\ref{eq:td1})
the integral along the trajectory $\psi(t)$ of the imaginary
component of the dot product between neighboring vectors along the
trajectory, and also to (\ref{eq:td2}) the integral of the average
phase velocity of the coefficients, weighted by the instantaneous
basis state probabilities $p_i(t) = r_i(t)^2$, which is
(\ref{eq:td3}) the time-integral of the instantaneous Hamiltonian
energy $E(t)=H(t)[\psi(t)]$ of the instantaneous state $\psi(t)$,
which (finally) is (\ref{eq:tdrel}) the integral of the
infinitesimal actions $\diff\alpha(t)=\langle
\psi(t)|A'(t)|\psi(t)\rangle$ on the instantaneous states $\psi(t)$.

The natural next question to ask is, given that $A'[\psi] =
\diff\alpha$ remains true over infinitesimal intervals $\diff t$ in
the general time-dependent case, and given that cumulatively,
$A(t)[\psi(0)] = \alpha$ in the time-independent case
($H(t)=H=\mathrm{const.}$), does this cumulative relation still hold
true in the general time-dependent case? That is, for $A(t)$ (as
defined in eq.~(\ref{eq:action})) is it still true that
\begin{equation}
    A(t)[\psi(0)] = \alpha  \label{eq:Aandalpha}
\end{equation}
even if the phase angle $\alpha$ was accumulated under the influence
of a varying $H(t)$?

If this equation~(\ref{eq:Aandalpha}) is universally correct, then
we will have a very nice, simple interpretation for the general
action operator $A(t)$ even in the case of a time-dependent $H(t)$,
namely that, when applied to any initial state $\psi(0)$, it simply
gives the angular length $\alpha$ of the trajectory that will be
traversed by that state, a quantity which obeys all of the
identities (\ref{eq:td1})-(\ref{eq:tdrel}).

Actually it seems that this is true, and the proof is quite elegant.
First, from eq.~(\ref{eq:contprod}) and the boundary condition
$U(0)=1$, fix $U=U(t)$, the overall unitary transform operating
between times 0 and $t$ that is implied by the values of the
time-dependent Hamiltonian $H(\tau)$ for all $0\leq\tau\leq t$. Fix
then also $A=A(t)$ by using eq.~(\ref{eq:logU}) and the associated
discussion, using the continuity requirement on $A(\tau)$ and the
requirement that $A(0)=0$.

Now, consider any eigenvector $|\phi_i\rangle$ of $U$, which is a state that
undergoes a cyclic evolution (in the projective Hilbert space) under
$H(\tau)$ or any other process (Hamiltonian trajectory) that
implements $U$, since $U|\phi\rangle=\mu_i|\phi_i\rangle$, with $\mu_i$ being the
associated unit-modulus eigenvalue.  Of course, $|\phi_i\rangle$ is then also an eigenvector
of $A$, with an eigenvalue $\alpha_i$ such that $A|\phi_i\rangle=\alpha_i|\phi_i\rangle$
and $\mu_i=\e^{\imag\alpha_i}$.

To see that this $\alpha_i$ must indeed be the same as the total
phase angle $\alpha$ accumulated by $\ket{\phi_i}$ as defined in
{\eg} eq.~(\ref{eq:tdrel}), consider that once the overall operator
$A$ has been determined, we can simply divide it by $t$ to find an
alternative {\emph{time-independent}} $H_c=A/t$ that would also
generate the very same action operator $A$ and the same unitary $U$
when applied over the same time interval $t$. From the discussion in
section 6, is is easy to see that the value of $\alpha$ is then
indeed exactly the phase angle accumulated from the initial state
$|\phi_i\rangle$ when implementing $A$ via this (alternative)
time-independent $H_c$.

Now, does {\emph{every}} Hamiltonian trajectory that implements $A$
(including our original time-dependent $H(\tau)$) involve the same
total accumulation $\alpha$ of phase angle?  We can see that it
must, because any trajectory $H(\tau)$ can, it seems, be
continuously deformed into the constant trajectory $H_c(\tau)=H_c$
while maintaining the same overall $A$ (and thus $U$) throughout the
deformation process.  At no point during this continuous deformation
process can the total phase $\alpha$ that is accumulated ever
change, since, to produce the same $U$, the total phase $\alpha$
must always remain congruent to $\alpha_i$ (mod $2\pi$), and it
would be impossible for the total phase accumulated to jump by a
multiple of $2\pi$ at any point during any continuous deformation of
the trajectory.

To see that this is true, recall from eq.~(\ref{eq:logU}) and the
associated discussion that any continuous $A(\tau)$ can be
characterized by a continuously varying eigenbasis $\{|u_i(\tau)\rangle\}$
of $U(\tau)$ (with a sort of $k$-dimensional continuous gauge
freedom, where $k$ is the Hilbert space dimension), and by implied
integer parameters $n_i(\tau)$ that select which of the logarithm
values must be used at each time point $\tau$.  As we continuously
deform the Hamiltonian trajectory $H(\tau)$ as well as the
eigenbases $\{|u_i(\tau)\rangle\}$ (and thus the gauges of the associated
eigenvalues $u_i(\tau)$), the set of time points $\tau$ at which the
$n_i(\tau)$ values change also changes continuously.  Nowhere during
this continuous, local process can the total angle $\alpha$ accumulated along the trajectory possibly
change discontinuously by a multiple of $2\pi$.

Thus, our arbitrary time-dependent $H(\tau)$ takes the eigenstate
$\ket{\phi_i}$ through the same total angle $\alpha$ as would the
constant $H_c$ for which we already know that $\langle
\phi_i|A|\phi_i\rangle = \alpha$.

The above discussion establishes that (regardless of the dynamics
$H(t)$) the $A$ operator that we derive from it always gives the
correct accumulated angle $\alpha$ for all eigenstates $\phi_i$ of
$A$; therefore it is also correct for arbitrary initial
superposition states $\psi(0)$ (and for mixed states as well).

For a final interesting observation, let $\alpha(\psi(0),t)$ denote
the angle $\alpha$ accumulated from the initial state
$|\psi(0)\rangle$ over time $t$, and note that since
\begin{equation}
\E{A(t)}{\psi(0)} = \alpha(\psi(0),t)
\end{equation}
for all initial $\psi(0)$, the time-derivative of the operator
$A(t)$ must satisfy
\begin{equation}
\E{\frac{\diff}{\diff t}A(t)}{\psi(0)}
    = \frac{\partial}{\partial t}\alpha(\psi(0),t).
\end{equation}
Recall meanwhile that $\diff\alpha(t)$ is given by applying
$A'(t)=H(t)\diff t$ to the state $\psi(t)$; \ie, $\diff\alpha(t) =
A'(t)[\psi(t)]$.  Of course, $\psi(t)=U(t)\psi(0)$, so we have that
\begin{eqnarray}
\langle\psi(0)|\frac{\diff A}{\diff t}(t)|\psi(0)\rangle
    &=& \frac{A'(t)}{\diff t}[U(t)\psi(0)] \\
    &=& \langle\psi(0)|U^\dagger(t)H(t)U(t)|\psi(0)\rangle.
\end{eqnarray}
and thus
\begin{eqnarray}
    \frac{\diff A}{\diff t}(t) &=& U^\dagger(t)H(t)U(t) \nonumber \\
                               &=& \e^{-\imag A(t)}H(t)\e^{\imag A(t)}.\label{eq:derivA}
\end{eqnarray}
Now, note that applying the time-dependent operator form
(\ref{eq:timedep}) of the Schr\"{o}dinger equation to
$U(t)=\e^{\imag A(t)}$, we get
\begin{eqnarray}
    \frac{\diff}{\diff t}\e^{\imag A(t)}
        &=& \imag H(t)\e^{\imag A(t)} \nonumber \\
        &=& \imag \e^{\imag A(t)}\e^{-\imag A(t)}H(t)\e^{\imag A(t)} \nonumber \\
        &=& \e^{\imag A(t)} \frac{\diff}{\diff t}[\imag A(t)],
\end{eqnarray}
where we have used~(\ref{eq:derivA}) in the last step.  In other
words, the ordinary rule $\diff\e^f=\e^f\diff f$ for the
differential of an exponential of a function $f$ actually turns out
to be true when $f=\imag A(t)$, despite the fact that the
Hamiltonian may be time-dependent and that $A(t)$ doesn't
necessarily even commute with its time-derivative! This is due to
the special way in which we defined our $A(t)$ function, and would
not be true for more general time-dependent operators.

\section{Discussion of Effort}

Although a choice of a particular cumulative action operator $A$
still gives us freedom to choose any number of different Hamiltonian
trajectories $H(\tau)$ for implementing it, over various total
amounts of time $t$, we have seen above that all such trajectories
are equivalent in terms of the total amount $\alpha$ of phase angle
that is accumulated starting from any fixed initial state
$\ket{\psi(0)}$.

As hinted previously, we might even consider the quantity $\alpha$
(or, more properly, its absolute value) to be a reasonable
definition of the geometric \emph{length} of the path that a
normalized state vector $\ket{\psi(t)}$ describes as it moves along
any continuous path (parameterized by any real variable $t$) along
the unit sphere in Hilbert space, since (note) $\alpha$ depends only
on the shape of the state trajectory itself, and not on any other
properties of the Hamiltonian trajectory, such as the energy of
other orthogonal states.

As a result, an intrinsic metric on the normalized Hilbert space is
provided by the distance function
\begin{equation}
d(\ket{\psi_1},\ket{\psi_2}) :\equiv \min |\alpha|
\end{equation}
where $\alpha$ is the accumulated phase angle along a given
trajectory, and the minimum is taken over all normalized, continuous
paths from $\ket{\psi_1}$ to $\ket{\psi_2}$, or a subset of such
that is deemed available.  The absolute-value operator is required
in order to obtain a proper (positive) metric, since trajectories
with unboundedly negative values of $\alpha$ could exist if we allow
states to have negative energy. Paths having the minimum absolute
$\alpha$ between a given pair of states can be considered to be
(sections of) geodesics on the normalized Hilbert space.

In \cite{Wootters-81}, Wootters introduced a statistically-motivated
distance metric between quantum states which he called ``statistical
distance,'' and showed that it was identical to the ordinary
Hilbert-space distance function $d(\psi_1,\psi_2)=\arccos
|\langle\psi_1|\psi_2\rangle|$.  It turns out that our distance
function $d$ above is in fact exactly the same as this also, if all
Hilbert-space trajectories are considered.  However, if the space of
allowed trajectories is restricted (for example, if the Hamiltonians
are forced to be local) then a different distance measure results.
In Wootters' metric, the distance between any two distinguishable
states (\eg, two different randomly chosen computational basis
states) is only $\arccos 0 = \pi/2$, whereas if we define distance
by minimizing over allowed trajectories, we could obtain a much
greater figure.

Later, we will see that our distance measure will also allow us to
derive a natural metric on unitary operations, telling us the
``distance'' between two unitaries, as measured by the difficulty of
getting from one to the other, in terms of the minimum distance
traversed by worst-case states.

Anyway, noting that this measure $\alpha$ of trajectory length which
we have explored above is stable with respect to changes of basis,
that there are multiple simple ways of defining it, and that it
connects strongly with fundamental physical concepts such as action
and energy, as well as with primitive geometric concepts such as
angles and areas, and that it forms a natural metric on the Hilbert
space, all of these facts together motivate us to propose this
measure as being the most natural and genuine measure of the total
``amount of change'' that is undergone by a physical quantum state
vector $\ket{\psi(t)}$ as it changes dynamically under a (possibly
varying) physical influence $H(t)$.

Insofar as we can consider {\emph{all}} dynamical evolution and
change to be forms of ``computation,'' where this word is construed
in a very general sense, we can also accept this measure as being an
appropriate measure of the \emph{amount of computational effort
exerted} by the system as it undergoes the given trajectory.

Thus, from here on, rather than calling our quantity ``action''
(which would lead to confusion with the action of the Lagrangian),
or ``accumulated phase angle'' (which is awkward) we will refer to
our quantity as simply the \emph{effort} when we wish to be concise,
and abbreviate it with the symbol $\eff$. That is,
\begin{equation}
    \eff_{t_1\rightarrow t_2}\left[\psi(t)\right]:\equiv \int_{t=t_1}^{t_2}\Im\langle
    \psi(t)|\psi'(t)\rangle
\end{equation}
is a real-valued functional of a state vector trajectory $\psi(t)$
taken between two times $t_1$ and $t_2$.  Note that the value of
$\eff$ depends only on the shape of the path.  It is independent of
the absolute time, the speed at which the trajectory is traversed,
and on various other details of the Hamiltonian that generates the
trajectory (such as its eigenvalues for eigenstates that are not
components of $\psi$); in general, many different Hamiltonian
evolutions can generate the same path, which will always have the
same total effort. So, in the above equation, we can consider
$\psi(t)$ to just be a parameterized curve where $t$ is now just any
\emph{arbitrary} real-valued parameter, not necessarily even
corresponding to physical time.  In other words, the \emph{effort}
quantity does not depend on the precise system of coordinates that
is used for measuring the passage of time, but rather only on a pure
geometric object, namely the path taken through Hilbert space.

Note that to say that the path length corresponds to computational
\emph{effort} is not to imply that all of the physical computation
that is occurring in the given system is necessarily being harnessed
and applied by humans to meet our calculational needs, only that
this is the total amount of raw computational work that is occurring
``in nature.''  The choice of the word ``effort'' is intended to
evoke the commonsense realization that effort may be wasted, \ie,
not used for anything useful.

Note also that the action operator $A$ (as we have defined it) gives
a concise yet particularly comprehensive characterization of a given
computational process, in the sense that it determines not only the
overall unitary operation $U=\e^{\imag A}$ that will be performed,
but also the amount of effort that will be expended in getting to
the final result from any given initial state.

The primary caveat to the above conception of computational effort
seems to be that the quantity $\eff$ (together with the rate of
phase rotation, and the path length in Hilbert space) is dependent
on where we choose to draw our zero of energy. As is well known,
absolute energies are only physically defined up to an additive
constant, and so the total Hamiltonian action or effort is only well
defined up to this constant multiplied by the elapsed time $t$.

A natural and widely-used convention is to define the least
eigenvalue of the Hamiltonian (the ``ground state'' energy) to be
the zero of energy.  In a similar fashion, we can choose to
additively shift the Hamiltonian so that the least eigenvalue of the
cumulative action operator $A(t)$ is taken to represent zero effort.
(Note that this approach can even be used when the Hamiltonian
itself is time-dependent.)

However, this choice is by no means mandated mathematically, and in
fact, in certain pathological cases (such as an infinite-dimensional
or time-dependent Hamiltonian with unboundedly negative
eigenvalues), there might not even be \emph{any} minimum eigenvalue
for the resulting action operator over a given interval. One needs
to keep these caveats in the back of one's mind, although they
seemingly end up not very much affecting the potential practical
applications of this concept, which we will address in a later
section.

Another reason that we might not want to consider the ground state
energy to always be zero is if the ground state energy varies,
especially if it includes energy that had to be explicitly
transferred into the system from some other external subsystem.
Thus, energy that is present in a given system, even if that system
is in its ground state, may still represent energy that was
transferred from elsewhere and isn't being used for other purposes;
\ie, it may represent ``wasted'' computational effort, and we may
wish to count it as such, rather than just counting it as zero
effort.

Another possible convention would be to count a system's energy as
being its total (gravitating) mass-energy, or rest mass-energy, if
we want it to be independent of the observer's velocity.  One might
think this choice is a somewhat less arbitrary than the ground state
convention, since mass is a physical observable, but unfortunately,
in general relativity, the contribution to the total mass-energy of
a local system that is due to its gravitational self-energy isn't
actually independent of the coordinate system that is used
(\cite{Dirac-75}, p.~62).  However, this caveat is usually only
important in extreme systems such as neutron stars and black holes,
where the gravitational self-energy contributes significantly to the
system's total mass.

In any case, for now, we propose to just make a ``gentlepersons'
agreement'' that we will always make sure that the energy
eigenvalues of the systems that we consider are always shifted so as
to be positive, so that the total effort is always positive, and we
don't have to worry about what would be the meaning of a negative
``amount of computational effort.'' Unfortunately, this strategy
rules out considering certain classes of systems, such as bottomless
potential wells, or the infinite Dirac sea of negative-energy
fermion states. But resolving this issue will have to wait for
future work.

\section{More Abstract Scenarios}
In the above, we have specified a well-defined (at least, up to an
additive constant) positive, real-valued measure $\eff$ of the
amount of computational effort represented by any trajectory of a
state vector in Hilbert space.

This raises the question of whether we can assign a measure of
computational effort to other physical situations that may be less
completely specified.  For example, we may be given a cumulative
action operator $A$, but not know the detailed Hamiltonian
trajectory $H(t)|_{t=t_1}^{t_2}$ that generated it, and we may be
given only a set $V$ of possible initial states (rather than a
single definite state), or we may have a probability distribution or
density function $p:V\rightarrow[0,1]$ over initial states.  In such
more abstract situations, can we still meaningfully define the
amount of computational effort exerted by the system as it undergoes
the evolution specified by its Hamiltonian over a given time
interval?

Of course we can.  Given a cumulative action operator $A$ and given
any specific state $\psi=\psi(t_1)$ at the initial time $t_1$, the
value of $\eff_{t_1\rightarrow t_2}[\psi(t)]$ is independent of the
details of the Hamiltonian trajectory $H(t)$ and is given simply by
\begin{equation}
\eff_A(\psi) :\equiv A[\psi] = \langle\psi|A|\psi\rangle,
\end{equation}
which can be called {\defn{the effort undergone by $\psi$ under
$A$}}.

We can therefore also naturally express the {\defn{average or
expected effort over $V$ exerted by the action operator $A$}} as:
\begin{equation}
\widehat{\eff}_V(A) = \mathrm{Ex}_V[\eff_A] = \sum_{\psi\in
V}p(\psi)\eff_A(\psi) = \langle A\rangle = \mathrm{Tr}(\rho A),
\end{equation}where the density
operator $\rho$ describing the initial mixed state is constructed
from the probability distribution over pure states $\psi$ in the
usual fashion, that is, with $\rho = \sum_{\psi\in
V}p(\psi)|\psi\rangle\langle\psi|$. If no probability distribution
$p$ has been provided, we can use a uniform distribution over some
natural measure on the set $V$.

This then gives us a workable definition of the mean effort exerted
by a system over time under a given Hamiltonian, even when the
initial state is not exactly known.

In some situations, we might also be particularly interested in the
\emph{maximum} effort over the set $V$ of possible initial states.
For example, suppose we are preparing the initial state of the
system, and we want to initialize the system in such a way that it
will exert the maximum effort possible.  Given $A$ and maximizing
over $V$, we define the {\defn{maximum effort exerted by $A$ over
$V$}} as
\begin{equation}
\eff_V^+(A) :\equiv \max_{\psi\in V} \eff_A(\psi).
\end{equation}
This can be considered to be a measure of the \emph{potential}
computational ``strength'' of the given action operator $A$,
expressing that any Hamiltonian $H(t)$ that implements $A$ over some
arbitrary interval $t_1\rightarrow t_2$ \emph{could} exert an amount
$\eff_V^+(A)$ of computational effort over that same interval, given
a suitable initial state. Insofar as the actual state that we end up
getting \emph{might} be the one that undergoes the maximal amount of
effort, we can say that a system with an unknown or unspecified
state is, at least, exerting this much ``potential'' computational
effort.

Even if the actual state turns out \emph{not} to be the
maximal-action one, the system could still be thought of as having
``done the work'' of determining that the actual state is \emph{not}
the one that should have transitioned through the given maximum
Hilbert-space distance. This particular thought should really be
credited to Seth Lloyd, who pointed out to me in personal
discussions, as an analogy, that an ordinary Boolean gate operation
can still be thought of as doing computational work even if the
output bit that it is applied to is not actually changed; namely, it
is doing the work of determining \emph{that} the bit should not
change.

Similarly to how we defined the maximum effort, we can likewise
define the {\defn{minimum effort of $A$ over $V$}} as $\eff_V^-(A)
:\equiv \min_{\psi\in V} \eff_A(\psi)$, although we should keep in
mind that if the ground state of the action operator $A$ is an
available initial state in $V$, and if we use the convention that
the ground state action is defined to be zero, then $\eff_V^-(A)$
will always be 0, and so will not be very useful.

\section{Difficulty of Performing an Operation}

Suppose now that we are given \emph{no} information about the
situation to be analyzed except for a unitary operator $U$ on the
Hilbert space $\mathcal{H}$, and we want to address the following
question: How much computational effort, at minimum, is required to
physically implement $U$?  By ``implement'' we mean that $U$ is the
time evolution operator that ends up being generated by the dynamics
over some interval, according to $U=\e^{\imag A}$ for some action
operator $A$.  We can call this minimum required effort the
{\emph{difficulty}} $\dif$ of implementing the unitary operator $U$.
Our framework gives us a natural way to formalize this notion.

Assuming we have some freedom of choice in the design of the system,
then among the set $\mathcal{A}$ of all Hermitian operators $A$ on
$\mathcal{H}$, or among at least a set $\aleph\subseteq\mathcal{A}$
of \emph{available} or implementable action operators, we might want
to choose the operator $A$ that generates $U$ that has the
\emph{smallest} value of the maximum or worst-case effort
$\eff_V^+(A)$ over the set $V$ of possible initial state vectors.
This $A$ can be considered to be the ``best'' action operator for
generating the given unitary $U$, in the sense that the length of
the longest trajectory that would be undergone by any possible state
vector $\psi\in V$ is minimized.  This strategy is analogous to what
we do in traditional algorithm design, where we usually choose the
algorithm that has the minimum time complexity on worst-case input
data. In our case, $A$ can be considered to abstractly represent the
algorithm selected, while the initial vector $\psi$ represents the
input data. Rather than time complexity, we focus on effort or
Hamiltonian action, since (as we will see) this translates directly
to time when a given supply of energy is available to be invested in
the system.

In some situations, it may be preferred to choose $A$ so as to
minimize the \emph{expected} effort rather than the worst-case
effort, for example, if we want to minimize the total effort exerted
over an arbitrarily large set of computations with randomly chosen
input states selected from some distribution.

We can thus define the maximum ($\dif_{\aleph,V}^+$) and expected
($\widehat{\dif}_{\aleph,V}$) difficulty of a desired unitary
transform $U$ under the available action set $\aleph$ and
initial-state set $V$ as follows:
\begin{eqnarray}
\dif_{\aleph,V}^+(U) &:\equiv& \min_{A\in\aleph} \eff_V^+(A) \nonumber \\
    &=& \min_{A\in\aleph} \max_{\psi\in V}\eff_A(\psi) \\
\widehat{\dif}_{\aleph,V}(U) &:\equiv& \min_{A\in\aleph} \widehat{\eff}_V(A) \nonumber \\
    &=& \min_{A\in\aleph} \sum_{\psi\in V}p(\psi)\eff_A(\psi)
\end{eqnarray}
Note that in all cases we still want to minimize over the available
action operators $A\in\aleph$, because there is usually no physical
reason why indefinitely large action operators (which waste
arbitrarily large amounts of effort) could not be constructed to
implement a given unitary; thus, maximizing over action operators
would thus always give $\infty$ and would not be meaningful.

A remark about the set $\aleph$ of available action operators.
Typically it would be constrained by what constitutes an
``available'' dynamics that we are free to choose within a given
theoretical, experimental, or manufacturing context.  For example,
$\aleph$ might reasonably be constrained to include only those
action operators that are obtainable from time-dependent
Hamiltonians $H(t)$ which are themselves constructed by summing over
local interaction terms between neighboring subsystems, or by
integrating a Hamiltonian density function that includes only local
terms on a field over some topological space, \eg, to reflect the
local structure of spacetime in a quantum field theory picture.  Or,
we might constrain ourselves to action operators that are obtainable
from time-independent Hamiltonians only, {\eg} if we are designing a
self-contained (closed) quantum system. Finally, practical
considerations may severely constrain the space of Hamiltonians to
ones that can be readily constructed in devices that can be built
using a specific manufacturing process, although we should note that
if scalable universal quantum computers can be built, then any
desired local Hamiltonian could be straightforwardly emulated on
these machines.

As a brief aside, it is also interesting to note that a given
difficulty function $\dif(U)$ (either the worst-case or average-case
version, and whatever $\aleph$ and $V$ are) also induces an
intrinsic metric on the space of unitaries of a given rank; we can
define a suitable distance function between unitaries by
\begin{equation}
d(U_1,U_2) = \dif(U_2U_1^\dag)
\end{equation}
that is, the distance between $U_1$ and $U_2$ in this metric is just
the difficulty of performing the relative unitary $U_{1\rightarrow
2}:\equiv U_2U_1^\dag$ that is equivalent to undoing $U_1$ (using
$U_1^\dag = U_1^{-1}$) and then doing $U_2$.  A unitary trajectory
for implementing $U_{1\rightarrow 2}$ that actually minimizes the
effort will then form, when right-multiplied by $U_1$, a (section of
a) geodesic in the space of unitaries passing between the unitaries
$U_1$ and $U_2$ (since $U_{1\rightarrow 2}U_1 = U_2$). Of course, in
general, the shortest unitary trajectory for implementing
$U_{1\rightarrow 2}$ will {\emph{not}} actually work by doing
$U_1^\dag$ followed by $U_2$; for example, if $U_1$ and $U_2$ have
high difficulty but are very close together, then the shortest
unitary trajectory between them will be much more direct than this.

Now, given our notion of the computational difficulty of a given
unitary $U$, we can now reinterpret previous results (such as
\cite{Margolus-Levitin-98,Levitin+02}) regarding ``quantum speed
limits'' or minimum times to implement various specific unitary
transforms of interest, or classes of transforms, given states of
specified average energy above the ground state, as follows:  These
analyses are implicitly specifying an $\aleph$ (usually, just all
Hermitian operators) and a $V$ (usually, just the entire Hilbert
space), and showing that the worst-case difficulty $\dif^+(U)$ for
the transform $U$ has a specific value (or lower bound), assuming
the presence of a time-independent Hamiltonian where the ground
state energy is usually set to 0.  In other words, such analyses
show that a certain minimum worst-case effort or Hamiltonian action
is required to implement the particular $U$ in question.

As an example, Margolus and Levitin's result
\cite{Margolus-Levitin-98} can be interpreted as telling us that any
$U$ that rotates some state $\psi$ to an orthogonal state has a
worst-case difficulty of $\dif^+(U)\geq h/4$, since their result
shows that any state of energy $E$ takes time at least $h/4E$ (no
matter what the Hamiltonian) to accumulate the action needed to take
it to an orthogonal state; thus the Hamiltonian action $A=Et$ that
is required to carry out such a transition is at least $h/4$.

Another result in \cite{Margolus-Levitin-98} implies that if there
is a $\psi$ such that ($\ket{\psi}$, $U\ket{\psi}$, $U^2\ket{\psi}$,
$\ldots$, $U^{N-1}\ket{\psi}$, $U^{N}\ket{\psi} = \ket{\psi}$)
comprises a cycle of $N$ states, with each orthogonal to the
preceding and succeeding states in the cycle, then
$\dif^+(U)\geq\frac{h}{2}\frac{N-1}{N}$, even if we are given
complete freedom in constructing the Hamiltonian, aside from a
requirement that it be time-independent. For $N=2$, this expression
reduces to $h/4$, while for $N\rightarrow\infty$, it goes to $h/2$.
Thus, any physical computation that proceeds autonomously though an
unbounded sequence of distinct states must exert at least $h/2$
effort per state transition.

Notice that the Margolus-Levitin theorem is, strictly speaking, only
giving us a \emph{lower bound} on the worst-case difficulty, since
it is considering only a particular state $\psi$ of interest
(namely, one that actually undergoes a transition to an orthogonal
state), rather than finding the worst-case potential effort to
perform the corresponding $U$, maximized over all possible initial
$\psi$ in the Hilbert space. Later, we will see that the actual
worst-case effort for an orthogonalizing transformation is actually
$h/2=\pi$ even in the $N=2$ case, and possibly even higher in cases
that go through more states.

We anticipate that, armed our definitions, it would be a highly
useful and worthwhile exercise to systematically go through a
variety of the quantum unitary transforms that have already been
identified in quantum computing as comprising useful ``quantum logic
gate'' operations, and quantify their worst-case and average
difficulty, according to the above definitions, under various
physically realistic sets of constraints. This would directly tell
us how much physical Hamiltonian action is required to carry out
those operations (given a best-case Hamiltonian implementation,
while operating on a worst-case or average-case input state).  We
can likewise do the same for classical reversible Boolean logic
operations embedded within unitary operations, as well as classical
irreversible Boolean logic operations embedded within classical
reversible operations, with ancilla bits used as needed for carrying
away garbage information to be discarded.

Such an investigation will, for the first time, give us a natural
and physically well-founded measure of the physical complexity of
logic operations, in terms of Hamiltonian action.  This in turn
would directly tell us the minimum physical time to perform these
operations within any physical system or subsystem using a set of
states having a given maximum energy about the ground state, given
the known or prespecified constraints on the system's initial state
and its available Hamiltonian dynamics.  This new quantification of
computational complexity may also allow us to derive lower bounds on
the number of quantum gates of a given type that would be required
to implement a given larger transformation in terms of smaller ones,
and possibly to show that certain constructions of larger gates out
of smaller ones are optimal.

In subsequent subsections, we begin carrying out the above-described
line of research, with some initial investigations of the difficulty
of various simple operations in situations where the available
dynamics is relatively unconstrained, which is the easiest case to
analyze.

\section{Specific Operations}\label{sec:specops}
In this section, we explore the difficulty (according to our
previous definitions) of a variety of important quantum and
classical logic operations.

We will begin by considering some educated guesses about the
difficulty of various unitaries.  For each unitary $U$ we are to
imagine implementing it via a particular transformation trajectory
$U'(t)$ (and Hamiltonian $H(t)$ such that $U'(t)=\e^{\imag H(t)\diff
t}$) that is as ``direct'' as possible, in the sense of minimizing
the Hilbert-space distance through which worst-case states are
transported.  Intuition tells us that these minimal trajectories are
expected to follow geodesics in the space of unitaries, as per the
metric we defined earlier; in other words, they should be
``straight-line'' paths, so to speak, that get us to the desired
unitary as directly as possible.

\subsection{General two-dimensional unitaries}
Let us begin by considering $\mathrm{U}_2$, the space of unitary
transformations on Hilbert spaces of dimensionality 2.  In quantum
computing, these correspond to single-qubit quantum logic gates.  As
is well known (\eg, see \cite{Nielsen-Chuang-00}, eq.~4.9), any such
$U$ can be decomposed as
\begin{equation}
U=\e^{\imag\alpha}R_{\hat{n}}(\theta)
\end{equation}
where $\hat{n}=(n_x,n_y,n_z)$ is a real 3D unit vector and
$R_{\hat{n}}(\theta)$ is a Bloch-sphere rotation about this vector
by an angle of $\theta$, that is,
\begin{equation}
R_{\hat{n}}(\theta) = \e^{\imag(\theta/2)(\hat{n}\cdot\vec{\sigma})}
\end{equation}
where $\vec{\sigma}=(\sigma_x, \sigma_y, \sigma_z)$ is the vector of
Pauli matrices
\begin{equation}
\sigma_x = \left[%
\begin{array}{rr}
  0\, & 1 \\
  1\, & 0 \\
\end{array}%
\right],
\sigma_y = \left[%
\begin{array}{rr}
  0 & -\imag \\
  \imag & 0 \\
\end{array}%
\right],
\sigma_z = \left[%
\begin{array}{rr}
  1 & 0 \\
  0 & -1 \\
\end{array}%
\right].
\end{equation}
Let us now consider breaking down $U$ into its multiplicative
factors $\e^{\imag\alpha}$ and $R_{\hat{n}}(\theta)$, which we
observe commute with each other, since $\e^{\imag\alpha}$ is a
scalar. Thus, we can consider these two components of $U$ to be
carried out in either order, or even simultaneously if we prefer.

Let's start by looking at $R_{\hat{n}}(\theta)$.  At first, we might
guess that the worst-case effort that is required to perform
$R_{\hat{n}}(\theta)$ for angles $\theta$ where
$-\pi\leq\theta\leq\pi$ ought to just turn out to be $|\theta|/2$,
since, for example, a Bloch sphere rotation through an angle of
$\theta=\pi$ radians corresponds to inverting a spin in ordinary 3D
space through an angle of $180^\circ$ to point in the opposite
direction, which is an orthogonalizing transformation, and we
already know from the Margolus-Levitin theorem that any transition
to an orthogonal state under a constant Hamiltonian requires a
minimum action (given zero ground state energy) for the state in
question of $h/4 = (\pi/2)\hbar = (\pi/2)\,\mathrm{rad}$, or an area
swept out of $\pi/4$ square units.  This is a good first guess, but
later, we will see that the actual worst-case action turns out to be
twice as large as this.  (Our intuition forgot to take into account
the fact that the state vector in the Margolus-Levitin theorem isn't
actually the worst-case one, as far as the accumulated Hamiltonian
action is concerned.)

Indeed, for any real unit 3-vector $\hat{n}$ (the ``axis of
rotation'' for the Bloch sphere), one can easily verify that there
is always a corresponding complex state vector
\begin{equation}
\ket{v_{\hat{n}}^+} =
\frac{1}{\sqrt{2(1+n_z)}}\left[\begin{array}{l}
                                                 n_z+1 \\
                                                 n_x+\imag n_y \\
                                               \end{array}\right]
\end{equation}
which is a unit eigenvector of $\hat{n}\cdot\vec{\sigma}$ having
eigenvalue +1.  This state vector is therefore also an eigenstate of
$R_{\hat{n}}(\theta)$, with eigenvalue $\e^{\imag(\theta/2)}$. In
other words, in any orthonormal basis that includes
$\ket{v_{\hat{n}}^+}$ as one of the basis vectors, as $\theta$
increases from 0 (for now, we'll assume for simplicity that the
final value of $\theta$ is non-negative, $0\leq\theta\leq\pi$), the
coefficient of the $\ket{v_{\hat{n}}^+}$ component of the state
$\ket{\psi(t)}=R_{\hat{n}}(\theta)\ket{v_{\hat{n}}^+}$ (starting
from the initial state $\ket{\psi(0)}=\ket{v_{\hat{n}}^+}$, where
the coefficient $c_{\ket{v_{\hat{n}}^+}}$ is 1) describes a circular
arc in the complex plane centered on the origin, sweeping out a
total angle of $\theta/2$, and an origin-centered area of
$\theta/4$.  As we saw earler, this same measure of the
weighted-average accumulated angle and total area accumulated still
holds in any basis.  So, we have that the effort of
$R_{\hat{n}}(\theta)$ must be at least $\theta/2$. Indeed, this is
the exact worst-case effort, since $\ket{v_{\hat{n}}^+}$'s
eigenvalue is maximal, so no pure energy eigenstate can possibly
sweep out a larger angle as $\theta$ increases, and therefore no
superposition of energy eigenstates (\ie, no general state) can do
so either.

Now, what about the $\e^{\imag\alpha}$ factor that's included in the
expression for a general $U\in\mathrm{U}_2$?  Note that this term
represents an overall (global) phase factor that applies to all
eigenstates.  As such, even the ground state $\ket{g}$ of whatever
Hamiltonian is used to implement $U$ might still accumulate a phase
due to this phase factor.  In this case, $\ket{g}$ would have
nonzero Hamiltonian energy.  If we redefine $\ket{g}$ to instead
have zero energy ($H\ket{g}=0$), then $\ket{g}$'s coefficient would
not phase-rotate at all, since the action operator $A=Ht$ would give
$A\ket{g}=0$ for this state, and $U\ket{g}$ would give $(\e^{\imag
A})\ket{g} = (\e^0)\ket{g} = \ket{g}$, that is, $\ket{g}$ would be
unchanged by this $U$. However, it does not follow that we can
always just let $\alpha$ be zero, as $\ket{g}$ may generally have
accumulated an additional phase resulting from the
$R_{\hat{n}}(\theta)$ component of $U$ as well. It is the
\emph{total} phase accumulated by the ground state that we wish to
define to be zero.

Let us now consider the following:  Under the transformation
$R_{\hat{n}}(\theta)$, as $\theta$ increases from 0, we notice that
$\ket{v_{\hat{n}}^+}$ (the eigenvalue-1 eigenstate of
$\hat{n}\cdot\vec{\sigma}$ which we constructed above) only
phase-rotates by an angle $\theta/2$.  Under
$U=\e^{\imag\alpha}R_{\hat{n}}(\theta)$, $\ket{v_{\hat{n}}^+}$
therefore undergoes an overall phase-rotation by an angle of
$\alpha+\theta/2$.  We confidently conjecture that the ``least
potential action'' or most efficient way to implement $U$ is to
apply a Hamiltonian that simultaneously sweeps both $\alpha$ and
$\theta$ forward steadily from 0, at respective rates that are
exactly proportional to their intended final values.  If this is
correct, then $\ket{v_{\hat{n}}^+}$ is indeed an eigenstate of that
best-case Hamiltonian, with energy $(\alpha+\theta/2)/t$ (recall
that we're using $\hbar=1$), where $t$ is the total time taken for
$\alpha$ and $\theta$ to reach their final values.

However, since the space we are working with is two-dimensional,
there must be another energy eigenstate as well.  Solving the
eigen-equation $(\hat{n}\cdot\vec{\sigma})\ket{v} = r\ket{v}$, we
find that the other eigenvalue $r$ of $\hat{n}\cdot\vec{\sigma}$ is
$-1$, and the other unit-length eigenvector, modulo phase-rotations,
is (for $n_z>0$)
\begin{equation}
\ket{v_{\hat{n}}^-} =
\frac{1}{\sqrt{2(1-n_z)}}\left[\begin{array}{l}
                                                 n_z-1 \\
                                                 n_x +\imag n_y \\
                                               \end{array}\right]
\end{equation}
or, in the special case when $n_z=0$, then instead any normalized
column vector $\ket{v_{\hat{n}}^-} = [v_0;v_1]$ where
$|v_0|=|v_1|=2^{-1/2}$ will work, so long as the vector components
$v_0$ and $v_1$ have the specific obtuse (that is, $>90^\circ$)
relative phase angle that is given by the relation $v_1 =
(-n_x-\imag n_y)v_0$.  (Note that $|n_x+\imag n_y|=1$ when $n_z=0$.)

Thus, for any Hamiltonian that smoothly sweeps $\theta$ forward in a
steady transformation $R_{\hat{n}}(\theta)$ with $\theta\propto t$,
there will actually be two different energy eigenstates having
energies that are negatives of each other, one state in which the
accumulated action of the Hamiltonian is $\theta/2$ (as we saw
above), and another state (the ground state) where the action is the
negative of this, or $-\theta/2$.  Together with the global
phase-rotation of $\alpha$, we have that the total action for $U$ is
$\alpha+\theta/2$ and $\alpha-\theta/2$ for these two energy
eigenstates, respectively.

Following our convention that the total action in the ground state
should be always considered to be zero, we can shift the energy
levels upwards in such a way that the lower value $\alpha-\theta/2$
will be equal to 0, in other words, we can adjust our rate of global
phase rotation (which determined $\alpha$) in such a way that we
have exactly $\alpha=\theta/2$.  Now, the total action in the high
energy state is $\alpha+\theta/2 = \theta/2 + \theta/2 = \theta$.

In other words, starting with any $U\in\mathrm{U}_2$ and decomposing
it as $U=\e^{\imag\alpha}R_{\hat{n}}(\theta)$, which involves a
rotation of the Bloch sphere through an angle of $\theta$ about an
axis $\hat{n}$, we can calculate a meaningful difficulty $\dif^+(U)$
by using the convention that the ground state should be considered
to have energy 0, and by letting $\dif^+(U) =
\dif^+(U_{\hat{n}}(\theta))$, where we define
$U_{\hat{n}}(\theta)\equiv\e^{\imag\theta/2}R_{\hat{n}}(\theta)$,
that is, ignoring the original value of $\alpha$ (whatever it was)
and instead adjusting $\alpha$ to have the value $\alpha=\theta/2$
which assigns the ground state to zero energy.  Thus, we can say
that the ``true'' computational/physical difficulty of $U$ (given
this choice) is exactly $\theta$ for any single-qubit unitary
$U=\e^{\imag\alpha}R_{\hat{n}}(\theta)$, regardless of the value of
$\alpha$.  If $\theta$ is a pure number (implicitly bearing an angle
unit of radians), then the worst-case Hamiltonian action to carry
out the desired transform using the best-case Hamiltonian (assuming
that is indeed what we have managed to characterize above) is
$\theta\hbar$, in whatever physical units we wish to express
$\hbar$.  That is, $\dif^+(U)=\theta$.

To wrap up this section, let us take a look at the precise form of
the Hamiltonian that we are proposing.  Note that
\begin{equation}
\hat{n}\cdot\vec{\sigma}=\left[\begin{array}{cc}
                                 n_z & n_x-\imag n_y \\
                                 n_x + \imag n_y & -n_z
                               \end{array}\right]
\end{equation}
is itself an Hermitian operator which plays the role of the
Hamiltonian operator $H$ with respect to the Bloch-sphere rotation
unitary
$R_{\hat{n}}(\theta)=\e^{\imag(\theta/2)(\hat{n}\cdot\vec{\sigma})}$,
if the rotation angle $\theta$ is taken be equal to twice the time
$t$.  Meanwhile, in this scenario, the extra phase-rotation factor
$\e^{\imag\alpha}=\e^{\imag(\theta/2)}$ out front corresponds simply
to an additional constant energy of +1, using the same angular
velocity units of $(\theta/2t)$.  This gives us a total
``Hamiltonian'' (in quotes because we haven't introduced an explicit
time parameter here yet) of $H_{\hat{n}}$ that is required to
implement a steady rotation about $\hat{n}$ which is equal to
\begin{eqnarray}
H_{\hat{n}} &=& 1 + \hat{n}\cdot\vec{\sigma} \nonumber \\
    &=& \left[\begin{array}{cc}
                1\,\, & 0 \\
                0\,\, & 1 \\
              \end{array}\right] +
        \left[\begin{array}{cc}
                n_z & n_x-\imag n_y \\
                n_x + \imag n_y & -n_z \\
              \end{array}\right] \nonumber \\
    &=& \left[\begin{array}{cc}
                1+n_z & n_x-\imag n_y \\
                n_x+\imag n_y & 1-n_z \\
              \end{array}\right].
\end{eqnarray}
With this choice of ``Hamiltonian,'' we can easily check that the
$\ket{v_{\hat{n}}^\pm}$ are indeed its energy eigenstates, with
$H_{\hat{n}}\ket{v_{\hat{n}}^-} = 0$ (the ground state has
``energy'' 0) and $H_{\hat{n}}\ket{v_{\hat{n}}^+} = 2$, which is
what we want since it will cancel out with the 2 in the denominator
of the exponent in the rotation unitary $U_{\hat{n}}(\theta) =
\e^{\imag\theta/2}R_{\hat{n}}(\theta) =
\e^{\imag(\theta/2)(1+\hat{n}\cdot\vec{\sigma})} =
\e^{\imag(\theta/2)H_{\hat{n}}}$.

To generalize the picture slightly, if a rotation through $\theta$
about an axis $\hat{n}$ is to take place over an arbitrary amount of
time $t$, then we require a Hamiltonian (a proper one now, in actual
angular-velocity energy units) of
\begin{equation}
H = \frac{\theta}{2t}H_{\hat{n}} = \frac{\theta}{2t}\left[
\begin{array}{cc}
                1+n_z & n_x-\imag n_y \\
                n_x+\imag n_y & 1-n_z \\
\end{array}\right]
\end{equation}

With this choice of Hamiltonian, note that things works out nicely
so that the high-energy eigenstate $\ket{v_{\hat{n}}^+}$
phase-rotates at exactly the desired rate $\omega^+ = \theta/t$,
since we have that
\begin{equation}
H\ket{v_{\hat{n}}^+} =
\frac{\theta}{2t}H_{\hat{n}}\ket{v_{\hat{n}}^+}
    = \frac{\theta}{2t}2\ket{v_{\hat{n}}^+} =
    \frac{\theta}{t}\ket{v_{\hat{n}}^+}
    = \omega^+\ket{v_{\hat{n}}^+}.
\end{equation}
Thus, the action operator $A=Ht$ comes out exactly equal to the
angle operator $\Omega$ which gives the total angle of phase
rotation for both the energy eigenstates $\ket{v_{\hat{n}}^\pm}$,
that is, $A\ket{v_{\hat{n}}^-} = \Omega \ket{v_{\hat{n}}^-} = 0
\ket{v_{\hat{n}}^-}$ and $A\ket{v_{\hat{n}}^+} = \Omega
\ket{v_{\hat{n}}^+} = \theta \ket{v_{\hat{n}}^+}$. And for an
arbitrary initial state $\psi$, \ie, for any normalized complex
superposition of the eigenstates $\ket{v_{\hat{n}}^\pm}$, $A[\psi] =
\Omega[\psi]$ gives the quantum mean angle of phase rotation.

% Of course, as usual, if we ever wish to convert from
% angular-velocity/angle units for energy/action (respectively) back
% to the usual kinds of physical units, we can simply multiply $H$ (or
% $A$) by $\hbar$, which makes the implicit angle unit ``1 radian''
% explicit in whatever arbitrary system of units we are using.

Note that in all the above discussion, we have assumed that the
rotation angle is non-negative, \ie, that $0\leq\theta\leq\pi$
(rad).  To complete the picture, note that for values of $\theta$
between 0 and $-\pi$, we can convert them to positive angles by the
simple expedient of rotating instead by an angle of $|\theta| =
-\theta$ about the $-\hat{n}$ axis , which is an exactly equivalent
rotation.  This has the effect of exchanging the values of the
$\ket{v_{\hat{n}}^\pm}$ eigenstates, as well as the sign of the
$H_{\hat{n}}$ component of $H$.  Other than that, everything else is
the same, with the result that the action $A$ always comes out
non-negative and equal to the absolute value of $\theta$.  Of
course, for the case of absolute angles outside the range
$(-\pi,\pi]$, we can just reduce them to the equivalent angle in
$(-\pi,\pi]$ by adding or subtracting the appropriate multiple of
$2\pi$.

In the above, although we have not yet quite finished proving
rigorously that the specific $H$ we have given is in fact the one
that implements $U$ with the least possible value of the worst-case
action $A$, still, we expect that it should already seem highly
plausible to the reader that this should in fact be the case, due to
the directness and simplicity of our construction, which made use
only of the simple fact that any arbitrary $U\in\mathrm{U}_2$ can be
decomposed into a single generalized rotation about an arbitrary
axis is real three-space, accompanied by a global phase rotation. Of
course, a more complete proof of the optimality of this construction
would be desirable to have, but it will have to wait for future
work.

\subsection{Specific single-qubit gates}
Given the above discussion, to determine the difficulty $\dif$ of
any single-qubit gate $U$ is a simple matter of finding some unit
3-vector $\hat{n}$ and angles $\alpha,\theta \in (-\pi,\pi]$ such
that $U=\e^{\imag\alpha}R_{\hat{n}}(\theta)$, which is always
possible. This then establishes that $\dif^+(U)=|\theta|$, under our
ground zero energy convention.  Let us look briefly at how this
calculation comes out for various single-qubit gates of interest.
\begin{enumerate}
    \item The Pauli spin-operator ``gates'' $X=\sigma_x$ (which is
    the in-place NOT operation in the computational basis),
    $Y=\sigma_y$, and $Z=\sigma_z$ all of course involve a rotation
    angle of $\theta=\pi$, since they all square to the identity
    ($2\pi$ rotation).  Thus, $\dif^+(X)=\dif^+(Y)=\dif^+(Z)=\pi=h/2$.
    \item The ``square root of NOT'' gate $N=\frac{1}{2}[\begin{array}{rr}
   \scriptstyle 1+\imag & \scriptstyle 1-\imag \\[-6pt]
   \scriptstyle 1-\imag & \scriptstyle 1+\imag
\end{array}]$ of course requires an angle of $\pi/2$, since $N^2=X$.
 Thus, $\dif^+(N)=\pi/2=h/4$.
    \item The Hadamard gate $N=\frac{1}{\sqrt{2}}[\begin{array}{rr}
  \scriptstyle 1 & \scriptstyle 1 \\[-6pt]
  \scriptstyle 1 & \scriptstyle -1
\end{array}]$ requires a rotation angle of $\pi$ about the
$\hat{n}=(1,0,1)/\sqrt{2}$ axis, \ie, $\hat{n}\cdot\vec{\sigma} =
(\sigma_x+\sigma_z)/\sqrt{2}$.  Also note that $H^2=1$ and a
rotation through $2\pi$ is the identity.  Thus, $\dif^+(H)=\pi=h/2$.
    \item The ``phase gate'' $S=[\begin{array}{rr}
  \scriptstyle 1 & \scriptstyle 0\\[-6pt]
  \scriptstyle 0 & \scriptstyle \imag
\end{array}]$ requires $\theta=\pi/2$ since note that $S^2=Z$.  So,
$\dif^+(S)=\pi/2=h/4$.
    \item The so-called ``$\pi/8$'' gate $T=[\begin{array}{cc}
  \scriptstyle 1 & \scriptstyle 0 \\[-6pt]
  \scriptstyle 0 & \scriptstyle \exp[\imag\pi/4]
\end{array}]$ involves $\theta=\pi/4$ since note that $T^4=Z$.
Thus, $\dif^+(T)=\pi/4=h/8$.
    \item The generalized phase gate $\mathrm{ph}(\theta)=[\begin{array}{cc}
 \scriptstyle 1 & \scriptstyle 0 \\[-6pt]
 \scriptstyle 0 & \scriptstyle \exp[\imag\theta]
\end{array}]$ is just a rotation by an angle of $\theta$ about the
$z$ axis, so $\dif^+(\mathrm{ph}(\theta)) =\theta = \theta\hbar$.
\end{enumerate}
As a point of comparison, the paper {\cite{Levitin+02}} studies the
time required to perform the specific gate $U=\e^{\imag\theta}X$
(\ie, NOT with global phase rotation) using an optimal Hamiltonian,
and conclude that the minimum time $\tau$ required (for a specific
initial state) is
\begin{equation}
\tau = \frac{h}{4E}\left(1+2\frac{\theta}{\pi}\right).
\end{equation}
Note that the corresponding Hamiltonian action $\alpha$ or effort
$\eff$ is
\begin{eqnarray}
\alpha = \eff = E\tau &=& \frac{h}{4}+2\frac{h}{4}\frac{\theta}{\pi}
\nonumber \\
    &=& \frac{\pi}{2}\hbar + \theta\hbar \nonumber \\
    &=& \frac{\pi}{2} + \theta\mathrm{\,\,\,\,(with}\,\,\hbar=1).\label{eq:toffact}
\end{eqnarray}
At first glance, this might appear to contradict our claim that the
difficulty of such a $U$ ought to be exactly $\pi$.  However, we
should keep two things in mind.  First, in {\cite{Levitin+02}},
Levitin {\etal} are concerned with the time to carry out $U$ in the
case of a specific subset of initial states which will actually
transition to an orthogonal state in the time $\tau$.  However,
these particular states are not the ``worst-case'' ones from our
perspective, and so they don't determine the maximum effort. Rather,
the particular states under consideration in their paper all have a
mean energy of only $\bar{E}=(E_1+E_2)/2$, where $E_1$ and $E_2$ are
the low and high energy eigenvalues of the ideal Hamiltonian,
respectively. Letting $E_1=0$ (our ground zero assumption), we have
that $E_2=2\bar{E}$.  Since $E_2$ has the highest energy available
given this spectrum, the $E_2$ energy eigenstate accumulates more
action over the time $\tau$ than any other possible state, in
particular, double that of states with energy $\bar{E}=E_2/2$, and
thus it is the $E_2$ state that determines the worst-case action,
which is twice that of {\cite{Levitin+02}}, or in other words
$A=\pi$. The term involving $\theta$ in~(\ref{eq:toffact}) drops out
entirely, since as we already saw earlier, global phase shifts are
irrelevant when considering total action, under our convention that
the ground state action is always defined to be zero.  Levitin
{\etal} don't make this adjustment, because they are assuming that
the Hamiltonian has already been arranged in advance to have a
desired energy scale. Thus, the global phase rotation by $\theta$
leads to an extra additive $\theta$ in their expression
(\ref{eq:toffact}) for the action.

\subsection{Difficulty of achieving infidelity}
A natural and widely-used measure of the degree of closeness or
similarity between two quantum states $u,v$ is the \defn{fidelity},
which is defined (for pure states) as $F(u,v)=|\langle u|v\rangle| =
|u^\dag v|$.  (See \cite{Nielsen-Chuang-00}.)  Note that if the
actual state of a system is $u$, and we measure it in a measurement
basis that includes $v$ as a basis vector, the square of the
fidelity $p=F^2$ gives the probability that the measurement operator
will project the state down to $v$, and that $v$ will be seen as the
``actual'' state. (This is a ``quantum jump'' or ``wavefunction
collapse'' event, or, in the many-worlds picture, it is the
subjectively experienced outcome when the state of the observer
becomes inextricably entangled with that of the system.) Likewise
with the roles of $u$ and $v$ reversed. Thus, only when $F=0$ are
the states $u$ and $v$ orthogonal.

We can also define a related quantity, the ``infidelity''
$\mathit{Inf}(u,v)\equiv\sqrt{1-p} = \sqrt{1-F^2}$.  The squared
infidelity between $u$ and $v$ is then just the probability $1-p$
that if the actual state is $u$, then it will \emph{not} be taken to
$v$ by a projective measurement (in a measurement basis that
includes $v$), and vice-versa.  In other words, if $v$ is some old
state of a system, and $u$ is its new state, the squared infidelity
between $u$ and $v$ is the probability that the answer to the
question ``Is the state different from $v$ yet?'' will be found to
be ``yes'' when this question is asked experimentally by a
measurement apparatus that compares the state with $v$.

Let us now explore the minimum effort that is required in order for
some of the possible state vectors of a system to attain a given
degree of infidelity (relative to their initial states), in the case
of two-dimensional Hilbert spaces.  Note that not all vectors will
achieve infidelity; in particular, the eigenvectors of any
time-independent Hamiltonian will always have 0 infidelity.

We start by recalling from earlier that any 2-dimensional unitary
can be considered a rotation of the Bloch sphere about some axis in
ordinary (real-valued) 3-D space.  Since a simple change of basis
suffices to transform any axis to any other, we can without loss of
generality presume a rotation about the $z$ axis, represented by
\begin{equation}
R_{\hat{z}}(\theta) = \left[\begin{array}{cc}
                              \e^{-\imag\theta/2} & 0 \\
                              0 & \e^{\imag\theta/2} \\
                            \end{array}\right].
\end{equation}
We saw earlier that the effort of any such rotation (under the
ground-zero convention) is always exactly $\theta$.  What initial
state will gain infidelity most rapidly under this transformation?
Until we figure this out, let us allow the initial state to be a
general unit vector $\ket{v}=[v_0; v_1] = v_0|\mathtt{0}\rangle +
v_1|\mathtt{1}\rangle$ in the basis
${|\mathtt{0}\rangle,|\mathtt{1}\rangle}$.  Then
$\ket{u}=R_{\hat{z}}(\theta)\ket{v}=[\e^{-\imag\theta/2}v_0;
\e^{\imag\theta/2}v_1]$ as a column vector of complex coefficients.
Now the fidelity between $v$ and $u$ is
\begin{eqnarray}
F(v,u) &=& \left|\langle v|u\rangle\right| = \left|\langle
                v|R_{\hat{z}}(\theta)|v\rangle\right| \nonumber \\
    &=& \left|v_0^*\e^{-\imag\theta/2}v_0 +
                 v_1^*\e^{\imag\theta/2}v_1\right|
            \nonumber \\
    &=& \left|\e^{-\imag\theta/2}|v_0|^2+
                \e^{\imag\theta/2}|v_1|^2\right|
            \nonumber \\
    &=& \left|\left[\cos\frac{\theta}{2}-\imag\sin\frac{\theta}{2}\right]|v_0|^2
                  + \left[\cos\frac{\theta}{2}+\imag\sin\frac{\theta}{2}\right]|v_1|^2\right|
            \nonumber \\
    &=& \left|\left(\cos\frac{\theta}{2}\right)(|v_0|^2+|v_1|^2)+
                        \imag\left(\sin\frac{\theta}{2}\right)(|v_1|^2 - |v_0|^2)\right| \nonumber \\
    &=& \left|\left(\cos\frac{\theta}{2}\right)+
                \imag\left(\sin\frac{\theta}{2}\right)(|v_1|^2-|v_0|^2)\right|.
\end{eqnarray}
where in the last line we have made use of the fact that
$|v_0|^2+|v_1|^2=1$ for a normalized $v$.  Now, $F^2$ is the sum of
the squared real and imaginary components of the expression inside
the outermost absolute-value delimiters $||$ above:
\begin{eqnarray}
[F(u,v)]^2 &=& \Im^2[\langle v|u\rangle] + \Re^2[\langle v|u\rangle] \nonumber \\
    &=& \cos^2\left(\frac{\theta}{2}\right) +
         \sin^2\left(\frac{\theta}{2}\right)\left(|v_1|^2-|v_0|^2\right)^2
    \nonumber \\
    &=& \cos^2\left(\frac{\theta}{2}\right) +
        \sin^2\left(\frac{\theta}{2}\right)\left(1 - 4|v_1|^2|v_0|^2\right)
        \nonumber \\
    &=& 1 - 4\sin^2\left(\frac{\theta}{2}\right)|v_1|^2|v_0|^2,
    \label{eq:fidel}
\end{eqnarray}
where in getting from the second to the third line, we have again
made use of the fact that $|v_0|^2+|v_1|^2=1$.  We can reassure
ourselves that the last line of (\ref{eq:fidel}) is always in the
range [0,1], since $|v_0|^2|v_1|^2\leq 1/4$ given that
$|v_0|^2+|v_1|^2=1$. Note also that the fidelity is minimized when
$|v_0|^2 = |v_1|^2 = \frac{1}{2}$, that is, when the two $z$-basis
states are in an equal superposition.  This is then the ``worst
case'' (worst in terms of ``least fidelity'') which we wish to focus
on.

So now, the infidelity $I=\mathit{Inf}(u,v)=\sqrt{1-F^2(u,v)}$ comes
out to be a reasonably simple expression:
\begin{eqnarray}
\mathit{Inf}(u,v) &=& \sqrt{1-[F(u,v)]^2} \nonumber \\
    &=& \sqrt{4\sin^2\left(\frac{\theta}{2}\right)|v_1|^2|v_0|^2} \\
    &=& 2\left(\sin\frac{\theta}{2}\right)|v_0||v_1|.
\end{eqnarray}
Note that for any given angle of rotation in $0<\theta<\pi/2$, the
infidelity is maximized when $|v_0|=|v_1|=1/\sqrt{2}$.  For such
$v$, we have $|v_0||v_1| = \frac{1}{2}$ and so
\begin{equation}
\mathit{Inf}(u,v)=\sin\frac{\theta}{2}.
\end{equation}
Thus, if we wish that some system initially in state $v$ should
achieve a desired degree $I$ of infidelity (relative to its initial
state) using a transformation of minimum effort, we must choose a
unitary transformation that is a rotation $R_{\hat{n}}(\theta)$
about an axis $\hat{n}$ that is ``perpendicular'' to $v$, and rotate
by an angle $\theta = 2\cdot\arcsin(I)$.  The Hamiltonian action
$\alpha$ accumulated by ``worst-case'' (that is, maximum-energy)
vectors under this transformation is (by definition) the difficulty
$\dif^+(R_{\hat{n}}(\theta))$ of that unitary, and is given by
$\alpha=2\cdot\arcsin(I)$.

However, the specific initial vector $v$ that we are dealing with
will not have the maximum energy $E$ (relative to ground) but rather
half of this, or $E/2$, since half of its probability mass will be
in the high-energy state, and half in the zero-energy ground state.
Therefore, $v$'s total Hamiltonian action (amount of change) along
its trajectory will instead be exactly $\alpha(v)=\arcsin(I)$, a
wonderfully simple expression.  This $\alpha$ is the effort exerted
by the specific state $v$ as it traverses a maximally efficient path
for achieving infidelity $I=\sin\alpha$.

So, for example, suppose we want to cause some given initial state
$v$ to transition to a new state that has only a probability of at
most $p=1/2$ of being confused with the initial state if it were
measured.  This is to say that the infidelity between the states
should be at least $I=\sqrt{1-p} =1/\sqrt{2}$, which requires the
state to traverse a trajectory that has a length of at least $\theta
= \arcsin(I) = \arcsin(1/\sqrt{2}) = \pi/4 = h/8$, which can be done
using a minimum-difficulty unitary transform whose worst-case effort
is twice as great as this, or $\pi/2=h/4$, meaning that the
worst-case (maximum-energy) states of the system would traverse a
trajectory of this (greater) length under an optimal implementation
of such a transformation.

Assuming that the actual given initial state in question is assigned
an average energy of only $E$ above the ground state, it will take
time at least $t=h/8E$ to carry out a unitary transformation on this
state that achieves a probability above $1/2$ of distinguishing it
from the resulting state; whereas, if we are given that the
\emph{maximum} energy state in the qubit spectrum has energy $E$,
then it will take time at least $t=h/4E$ to carry out the transform.

In other words, to carry out an operation in time $t$ that yields a
50\% probability (or less) of conflation of some initial states with
their successors requires that the initial states in question must
have energy at least $E=h/8t$, and that states of energy at least
$E=h/4t$ must exist in the spectrum.

Note that the above results are also perfectly consistent with the
Margolus-Levitin theorem {\cite{Margolus-Levitin-98}}.  That is,
plugging in an infidelity of $I=1$ to represent a transition to an
orthogonal state, we find that the specific initial state's effort
$\eff(v) =\arcsin(1)=\pi/2$ while the worst-case difficulty for this
transform is $\theta=2\arcsin(1)=\pi$; these figures are twice that
for the previous example.  And so for a state to attain a 0\%
probability of conflation (\ie, to reach an orthogonal state)
requires that it have at least twice the energy as the previous
scenario, or $E=\pi/2t=h/4t$ (under the Hamiltonian used to carry
out the transformation), while other energy levels of at least
$\pi/t=h/2t$ must be present in the spectrum of the Hamiltonian
operator being used.

\subsection{Higher-dimensional operations}\label{highd}
Naturally, we are interested not only in unitaries in
$\mathrm{U}_2$, but also in higher dimensions, in particular,
unitaries in the groups $\mathrm{U}_{2^n}$, which correspond to
general ``quantum logic gate'' operations (really, arbitrary quantum
computations) operating on sets of $n$ qubits.

In particular, let us focus on the ``controlled-$U$'' gates with one
target bit, which take the general form (modulo qubit reorderings)
\begin{equation}
U' = \mathrm{C}^{n-1}U \equiv \left[
    \begin{array}{cccc}
      1 &  &  &  \\
       & 1 &  &  \\
       &  & \ddots &  \\
       &  &  & U \\
    \end{array}
\right]
\end{equation}
where we have $2^{n}-2$ ones along the diagonal, and a rank-2
unitary matrix $U$ in the lower-right corner.  In other words, for
computational basis states $|b_0b_1\ldots b_{n-1}\rangle$, whenever
the first $n-1$ qubits $b_0b_1\ldots b_{n-2}$ are not all 1's, the
state remains unchanged; otherwise, the unitary $U$ is performed on
the final qubit $b_{n-1}$.

We observe immediately that $\dif^+(U')\geq \dif^+(U)$, since all
the input states that undergo any change at all will undergo the
exact same transformation (in the subspace associated with the last
qubit) that they would if $U$ were just applied unconditionally.
Thus, the worst-case trajectories when conditionally applying $U$
can be no shorter than the worst-case unconditional trajectories
(under an optimal implementation).

Furthermore, if $U$ by itself would be optimally implemented by the
Hamiltonian $H$, then it is easy to believe that $U'$ would likewise
be optimally implemented by the Hamiltonian
\begin{equation}
H' = \left[
    \begin{array}{cccc}
      0 &  &  &  \\
       & 0 &  &  \\
       &  & \ddots &  \\
       &  &  & H \\
    \end{array}\right]
\end{equation}
that is, with 0's everywhere except for a copy of $H$ in the
lower-right $2\times 2$ submatrix.  It is easy to verify that this
$H'$, when exponentiated, indeed produces the desired $U'$. And
since its worst-case difficulty is equal to our lower bound
$\dif^+(U)$, it is in fact an optimal $H'$, assuming our earlier
conjecture about the optimality of $H$ is correct. In this case, if
$H'$ is actually an available Hamiltonian in the context one is
considering, then the effort of $U'$ is indeed exactly the same as
the effort of $U$.

We can see from this example that when we consider the full space of
mathematically describable Hamiltonians, we are likely to greatly
underestimate the effort, compared to what can actually be
implemented.  The typical known implementations of $U$ in terms of
small local quantum gates would require a number of orthogonalizing
operations that is at least linear in $n$, whereas in our case
above, the effort is constant (upper-bounded by $\pi$).  It seems
likely that the effort for a physically realistic ({\eg}
field-theory based) Hamiltonian for this class of $U$s would have to
be more than constant, since the interaction of $n$ qubits to
determine an outcome would appear to necessarily be a non-local
process.

In most physical situations of interest, we will not necessarily
have available Hamiltonians that are of any form desired, such as
the form $H'$ suggested above.  Instead, we may only have available
a more limited, perhaps parameterized suite of Hamiltonians, perhaps
ones that are formed by a sum or time-sequence of specific,
controllable, localized couplings having (say) at most 2 qubits
each, as is popularly represented in the quantum computing
literature using the schematic notation of quantum logic networks.

Obviously, whenever our space of available Hamiltonians is more
restricted than the simple ``all Hermitian operations'' scenario
analyzed above, the resulting values of $\dif^+(U)$ will in general
become much larger, and probably also much more difficult for us to
analytically calculate.  To compute $\dif^+(U)$ for Hamiltonians
that can plausibly be constructed within the context of particular
experimental frameworks that are readily physically realizable in
the lab (or in a manufactured product, \eg, a
someday-hopefully-to-be-realized commercial quantum computer) is
clearly a much more complex and difficult task than we have
attempted to tackle in this paper.  To address this problem more
fully will have to wait for future work.

Still, we hope that the present work can at least serve as a
fruitful conceptual foundation on which we can proceed to build
meaningful analytical and/or numerical analyses of the
physical/computational ``difficulty'' of performing various quantum
operations.  We also hope that this work will serve as a helpful
stepping stone for future investigators who wish to continue
exploring the many deep and rich interconnections between physical
and computational concepts.

\subsection{Classical reversible and irreversible Boolean operations}
\label{sec:classic}

Although in the above discussion we have focused on the effort
required to carry out quantum gate operations, it is easy to extend
the results to classical logic operations as well. Any classical
reversible operation is just a special case of a quantum gate where
the matrix elements of the unitary operator (in the computational
basis) are 0 or 1. For example, a reversible Toffoli gate or
Controlled-Controlled-NOT (CCNOT) is a special case of the
$\mathrm{C}^2U$ gate addressed in \S\ref{highd} above. Specifically,
since the $U$ in question is $X$ (NOT), which has a rotation angle
of $\pi$, the effort required for Toffoli must be at least $\pi$,
and indeed is exactly $\pi$ if arbitrary Hamiltonians can be
constructed.  Toffoli is a universal gate for classical reversible
computation, so a construction of any classical reversible circuit
out of Toffoli gates sets an upper bound (as a multiple of $\pi$) on
the difficulty of that computation, apart from any extra effort that
may be required to control transitions between gates (which could be
substantial, but is probably close to linear in the number of
operations performed).

As for ordinary irreversible Boolean operations, these can be
embedded into reversible operations as follows.  Consider, for
example, a standard boolean inverter, whose function is irreversible
as it is normally specified in an electrical engineering context.
The explicit function of an inverter is to destructively overwrite
its output node with the logical complement of its input. (Please
note that this function is distinct from that of a classical
reversible NOT operation, which simply toggles a bit in-place.) Due
to Landauer's principle, the physical information contained in the
output node cannot actually be destroyed, but is instead transferred
to reside in the environment. So, we can model the ordinary
inverter's function as a sequence of reversible operations as
follows:
\begin{enumerate}
    \item Exchange output bit with an empty bit in the device's environment
    \item Increment an ``environment pointer'' to refer to the next empty bit in some
    unbounded list
    \item Perform a CNOT between input node and (now empty) output node
\end{enumerate}
The first step can be understood as the emission from the device of
the old stored value of the bit, in the form of entropy.  The second
step can be viewed as implementing the continuous flow of entropy
away from the device, to make room for discarding the results of
subsequent inverter operations.  Finally, the third step carries out
the desired logical function.  The above breakdown is not
necessarily the simplest possible implementation of the classical
inverter (although it is probably close), but it at least sets an
upper limit on the number of quantum operations that are absolutely
required.

The first step can be carried out by a unitary SWAP operation
between the two bits in question.  The second step can be carried
out by an annihilate/create pair of operations that moves a
``particle'' by one position to point to the next empty location in
the environment; this corresponds to a unitary operation that
increments the state vector $|i\rangle$ of some subsystem that
specifies the integer location $i$ of the environment pointer.
Finally, the third step is just an ordinary CNOT, with an effort of
$\pi$. In principle, we could calculate and add up the effort for
all these steps, together with the effort needed to update a part of
the machine state that keeps track of which step we are on, to
arrive at an upper bound on the effort required to implement a
classical inverter operation. However, this calculation might not be
very meaningful unless we did more work to specify a detailed
physical setup that would allow us to confirm that such a bound was
achievable in a practical hardware implementation.

\section{Relation to Berry phase}
An interesting question to ask about our quantity $\eff$ is what
relationship (if any) it has to the classic notion of the geometric
or Berry phase of a quantum trajectory
\cite{Berry-84,Simon-83,Aharonov-Anandan-87,Anandan-Aharonov-88,Samuel-Bhandari-88,Anandan-Aharonov-90,Bose-DuttaRoy-91,Zeng-Lei-95}.
So far, the relationships between these concepts are not completely
clear, and working them out in more detail will have to wait for
future work. However, some initial remarks are in order.

Let $H(t)$ be any time-dependent Hamiltonian that implements the
unitary $U$ for $t$ going from 0 to $\tau$, and let $\ket{\psi}$ be
an eigenvector of $U$, with eigenvalue $\e^{\imag\phi}$.  The state
$\ket{\psi}$ thus undergoes a cyclic evolution in the projective
(phase-free) Hilbert space.  Aharonov and Anandan
{\cite{Aharonov-Anandan-87}} point out the relation $-\phi = \alpha
- \beta$ (the integrated form of their equation (2)), where $\alpha$
is the integral of the instantaneous Hamiltonian energy of the
state,
\begin{equation}
\alpha = \frac{1}{\hbar}\int_{t=0}^\tau \langle
\psi(t)|H(t)|\psi(t)\rangle \diff t\label{eq:AandA}
\end{equation}
and $\beta$ is a term given by
\begin{equation}
\beta = \int_{t=0}^\tau \langle
\tilde{\psi}(t)|\imag\frac{\diff}{\diff
t}|\tilde{\psi}(t)\rangle\diff t,
\end{equation}
where $\tilde{\psi}(t)$ is any continuously gauge-twiddled version
of $\psi(t)$ such that $\tilde{\psi}(0)=\tilde{\psi}(\tau)=\psi(0)$.
Aharonov and Anandan's paper {\cite{Aharonov-Anandan-87}} revolves
around their claim that this $\beta$ quantity is a generalized
version of the Berry phase that applies even to non-adiabatic
evolutions.

However, if the results of the present paper are correct, then
Aharonov and Anandan's $\beta$ is always an arbitrary value
congruent to 0 (modulo $2\pi$) and thus is not a physically
meaningful quantity. The reason is that the $\alpha$
in~(\ref{eq:AandA}) is exactly our $\alpha = A[\psi(0)]$, where
$U=\e^{-\imag A}$ (in the usual sign convention, which A\&A are
using), and thus $\psi(0)$ is also an eigenvector of $A$ with
eigenvalue $\alpha$, so $\ket{\psi(\tau)} = U\ket{\psi(0)} =
\e^{-\imag\alpha}\ket{\psi(0)}$. Since we are already given that
$\psi(\tau) = \e^{\imag\phi}\psi(0)$, it follows that $\phi \equiv
-\alpha$ (mod $2\pi$); thus $\beta\equiv 0$ (mod $2\pi$).  Any
desired multiple of $2\pi$ can always be selected for $\beta$ by
appropriate choice of the function $\tilde{\psi}(t)$. So, $\beta$
does not contain any information at all about the specific evolution
$\psi(t)$, and thus it is not a physically meaningful quantity.

It it interesting to note that the A\&A paper
{\cite{Aharonov-Anandan-87}} never actually shows that their
quantity $\beta$ can ever be different from 0 (mod $2\pi$), although
they do prove that $\beta$ has some other ``interesting'' properties
(such as being independent of the gauge of the original trajectory)
which of course are true trivially if $\beta$ is always congruent to
zero.

Thus, it seems that one implication of our results (assuming they
are correct) is that Aharonov and Anandan's particular version (at
least) of the ``geometric phase'' is a chimera, and does not really
exist. Further study is needed to verify this conclusion more
rigorously, and also to determine whether other definitions of the
Berry phase might escape from it, and retain a useful physical
meaning that relates in some way to our quantity $\alpha$.  Since
many researchers have reported the experimental detection of
Berry-type phases (\eg, see \cite{Falci-00}), it seems highly
unlikely that our results will turn out to nullify all versions of
the geometric phase for all quantum evolutions. However, as of this
writing, the correct resolution of the apparent discrepancy between
theory and experiment on this question is not yet clear.

\section{Conclusion}
In this paper, we have shown that any continuous trajectory of a
normalized state vector can be measured by a real-valued quantity
which we call the {\emph{effort}} $\eff$, which is given by the line
integral, along the trajectory, of the imaginary component of the
inner product between adjacent states along the trajectory. This
quantity is basis-independent, and is numerically equal to the
probability-weighted average phase angle accumulated by the basis
state coefficients (in radians), and to twice the area swept out by
the coefficients in the complex plane, and also to the action of the
time-dependent Hamiltonian along the trajectory, in units of
$\hbar$. This notion of effort can be easily extended to apply also
to transformation trajectories $U'(t)$ over time, as well as to an
overall resulting unitary transform $U$, where it measures the
difficulty $\dif$ or minimum effort (over available trajectories)
required to implement the desired transform in the worst case
(maximizing over the possible initial states). Our framework can be
used to easily rederive a variety of related results obtained by
earlier papers for various more specialized cases.

The major implication of these results is that there is indeed a
very definite sense in which we can say that the physical concept of
energy does indeed precisely correspond to the computational concept
of the rate of computation, that is, we can validly say that energy
\emph{is} the rate of physical computing activity, defined as the
rate of change of the state vector, according to the measure that we
have described in this paper. Furthermore, we can validly say that
physical action \emph{is} (an amount of) computation, defined as the
total amount of change of the state vector, in the sense we have
defined.

What about different specific types of energy, and specific types of
action? Later papers along this line of research will survey how
different types of energy and action can validly be identified with
computational activity that is engaged in different types of
processes.  For example, heat may be identified with energy whose
detailed configuration information is unknown (is entropy), rest
mass-energy can be identified with energy that is engaged in
updating a system's internal state in its rest frame, potential
energy with phase rotation due to emission/absorption of virtual
particles, and so forth.  As a preview, it turns out that we can
even make our computational interpretation consistent with special
relativity by subdividing the energy of a moving body (in a given
observer frame) into the \emph{functional} energy $\Phi$ that is
associated with updating the body's internal state (this turns out
to be just the negative Lagrangian $-L=H-pv$) and a \emph{motional}
part $M=pv$ (related to but not quite the same as kinetic energy)
that is associated with conveying the body through space;
relativistic momentum then turns out to be the motional
computational effort exerted per unit distance traversed.  Future
papers will elaborate on these related themes in more depth.

It is hoped that the long-term outcome of this line of thought will
be to eventually show how \emph{all} physical concepts and
quantities can be rigorously understood in a well-defined
mathematical framework that is also simultaneously well-suited for
describing physical implementations of desired computational
processes.  That is, we seek an eventual unifying mathematical
foundation that is appropriate for not only physical science, but
also for device-level computer engineering and for physics-based
computer science.  We expect that such a unifying perspective should
greatly facilitate the future design and development of maximally
efficient computers constructed from nanoscale (and perhaps,
someday, even smaller) components, machines that attempt to harness
the underlying computational resources provided by physics in the
most efficient possible fashion.

\bibliography{Frank-QIP-05-revised}
\bibliographystyle{unsrt}       % Use citation order.

%\end{article}
\end{document}